\pdfoutput=1
\documentclass[letterpaper,twocolumn,english,aps,showpacs,showkeys,preprintnumbers,prd,floatfix,nofootinbib,superscriptaddress]{revtex4-1}

\usepackage{amsmath,amssymb,bbm,color}
\usepackage{graphicx}
\usepackage{epsfig}
\usepackage{euscript}
\usepackage{pslatex}
\usepackage{hyperref}
\usepackage{xcolor}
\usepackage{fancybox}
\usepackage{enumerate}

\newcommand{\Meu}{\EuScript{M}}
\newcommand{\Mmf}{\mathfrak{M}}

\newcommand{\KK}{${\cal KK}$}
\def\Ordex#1{${\cal O}(#1)_{exp}$}
\def\Order#1{${\cal O}(#1)$}
\def\OrderLL#1{${\cal O}(#1)_{\rm LL}$}
\def\bbeta{\bar{\beta}}
\newcommand{\sfac}{\mathfrak{s}}
\def\hbeta{\hat{\beta}}

\def\st{\hbox{}} 

\begin{document}

\preprint{
\vbox{
\null \vspace{0.3in}
\hbox{BU-HEPP-13-03}
\hbox{IFJPAN-IV-2013-13}
}
}

\title{\null \vspace{0.3in}
KK MC 4.22: CEEX EW Corrections for $f\bar{f}\rightarrow f'\bar{f'}$
at LHC and Muon Colliders}

\author{S.\ Jadach}
\affiliation{Baylor University, Waco, TX, USA}
\affiliation{Institute of Nuclear Physics, Polish Academy of Sciences,\\
  ul.\ Radzikowskiego 152, 31-342 Cracow, Poland}

\author{B.F.L~Ward}
\affiliation{Baylor University, Waco, TX, USA}

\author{Z.~W\c{a}s}
\affiliation{Institute of Nuclear Physics, Polish Academy of Sciences,\\
  ul.\ Radzikowskiego 152, 31-342 Cracow, Poland}


\begin{abstract}
We present the upgrade of the coherent exclusive (CEEX) exponentiation
realization of the Yennie-Frautschi-Suura (YFS) theory used in our Monte Carlo (${\cal KK}$ MC) to the processes $f\bar{f}\rightarrow f'\bar{f'}, f=\mu,\tau,q,\nu_\ell , f'=e,\mu,\tau,q,\nu_\ell , q=u,d,s,c,b,t, \ell=e,\mu,\tau $ 
with $f\ne f'$, with an eye toward the precision physics 
of the LHC and possible high energy muon colliders.
We give a brief summary of the CEEX theory in comparison 
to the older (EEX) exclusive exponentiation theory
and illustrate theoretical results relevant to the LHC 
and possible muon collider physics programs.
\end{abstract}

\maketitle
\tableofcontents{}

\section{Introduction}

Given that the era of precision QCD at the LHC is upon us,
by which we mean theoretical precision tags at or below 1\% in QCD
corrections to LHC physical processes, computation of higher order EW corrections are also required: in the single $Z$ production process at the LHC for example,
a u quark anti-u quark annihilation hard process at the Z pole has a radiation probability strength factor of $\frac{4}{9}\frac{2\alpha}{\pi}\left(\ln(M_Z^2/m_u^2)-1\right)\cong 0.038$ if we use the value $m_u\cong 5.0$ MeV, the current quark mass value -- we return to the best choice for the quark masses below. Evidently, we have to take these EW effects into account at the per mille level if we do not wish that they spoil the sub-1\% precision QCD we seek in LHC precision QCD studies~\cite{bflw1}. Indeed, when the cut on the respective energy of the emitted photons is at $v_{min}$ in units of the reduced cms effective beam energy, the $0.038$ strength factor above is enhanced to 
$0.038\ln(1/v_{min})$ and can easily become ${\cal O}(1)$. This means we have to use resummation, realized by MC event generator methods, of the type we have pioneered in Refs.~\cite{eex} to make contact with observation based on arbitrary cuts in any precise way. We call the reader's attention here to the approaches of Refs.~\cite{dima,hor,fewz,den-ditt,wac} to EW corrections to such heavy gauge boson production at the LHC. It is well-known from LEP studies~\cite{lepewwg} that using only the exact ${\cal O}(\alpha)$ EW corrections is inadequate for per mille level accuracy on these corrections. Our studies below will
show that this is still the case. This means that the approaches 
in Ref.~\cite{dima,fewz,wac,den-ditt} must be extended to higher orders for precision LHC studies. We comment further below on the relation of our approach to that in Ref.~\cite{hor} as well\footnote{We remind the reader that, as it is done in Ref.~\cite{hor} for example, in the hadron collider environment, one can also use DGLAP-CS~\cite{dglap,cs} theory for the large QED corrections in the ISR, so that standard factorization methods are used to remove the big QED logs from the reduced hard cross sections and they occur in the solution of the QED evolution equations for the PDF's which can be solved from the quark mass $m_q$ to the factorization scale $Q\simeq M_Z$ here because QED is an infrared free theory; in what follows, we argue that we improve on the treatment of such effects with resummation methods we discuss presently.}.\par
Presently, we recall that in the case of single 
Z/$\gamma*$ production in high energy $e^+e^-$ annihilation our state of the art realization of such resummation is the CEEX YFS~\cite{yfs,ceex1} exponentiation we have realized by MC methods in the ${\cal KK}$ MC\footnote{The name ${\cal KK}$ MC derives from the fact that the program was published in the last year of the second millenium, where we note that K is the first letter of the Greek word Kilo, and from the fact that two of us (S.J. and Z.W.) were located in
Krakow, Poland and the other of us (B.F.L.W.) was located in Knoxville, TN, USA at the inception of the code.} in Ref.~\cite{kkmc}. We conclude that we therefore need to extend the incoming states that the ${\cal KK}$ MC allows to include the incoming quarks and anti-quarks in the protons colliding at the LHC. Previous versions of ${\cal KK}$ MC even though not adapted for the LHC were already found useful in estimations of theoretical systematic errors of other calculations~\cite{zwad1,zwad2}. We denote the new version of ${\cal KK}$ MC by version number 4.22, 
${\cal KK}$ MC 4.22. Our aims in the current discussion 
in its regard are to summarize briefly on
the main features of YFS/CEEX exponentiation~\cite{ceex1,ceex2} in the SM EW theory, 
as this newer realization of the YFS theory is not a generally familiar one, to discuss the changes required to extend the incoming beam choices in the ${\cal KK}$ MC from the
original $e^+e^-$ incoming state in Ref.~\cite{kkmc} to the more inclusive choices
$f\bar{f},\; f=e,\mu,\tau, q, \nu_\ell ,\; q=u,d,s,b,\; \ell= e,\mu,\tau$, and
to present examples of theoretical results
relevant for the LHC and possible muon collider~\cite{muclldr} 
precision physics programs.
For example, the muon collider physics program involves precision studies 
of the properties of the recently discovered BEH boson~\cite{beh} candidate~\cite{atlas1,cms1} and treatment of the effects of higher order EW corrections will be essential to the success of the program, as we illustrate below. 
\par

In the next section, we review the older EEX exclusive realization 
and summarize the newer CEEX exclusive realization of the
YFS~\cite{yfs} resummation in the SM EW theory; for, the YFS resummation is not generally familiar so that our review of the material in Refs.~\cite{eex,ceex1,ceex2} will aid the unfamiliar reader to follow the current discussion. 
We do this in the context of
$e^+e^-$ annihilation physics programs for definiteness for historical reasons.
In this way we illustrate the latter's 
advantages over the former, which is also very successful. 
We also stress the key common aspects of our MC implementations of 
the two approaches to exponentiation,
such as the exact treatment of phase space in both cases, 
the strict realization of the factorization theorem, etc. We stress that both of the realizations of YFS exponentiation are available in the ${\cal KK}$ MC 4.22 where both allow for the new incoming beams choices. This gives us important cross-check avenues required to establish the final precision tag of our results.
In Sect. 3, we discuss and illustrate the extension of the 
choices of the incoming beams in the ${\cal KK}$ MC realization
of CEEX/EEX. We illustrate results which
quantify the size of the EW higher order corrections in LHC and muon collider physics scenarios. Specific realizations of the results we present here
in the context of a parton shower environment will appear 
elsewhere~\cite{elsewh}.
Sect. 4 contains our summary. Appendix 1 contains a sample output.

\section{Review of Standard Model calculations for 
$e^+ e^-$  annihilation with YFS exponentiation}

There are many examples of
successful applications~\cite{eex} of our approach to 
the MC realization of the YFS
theory of exponentiation for $e^+e^-$ annihilation physics: (1), for 
$e^+e^- \to f\bar{f} +n\gamma$, $f=\tau,\mu,d,u,s,c$ there are
YFS1 (1987-1989) \Ordex{\alpha^1} ISR,
YFS2$\in$KORALZ (1989-1990), \Ordex{\alpha^1+h.o.LL} ISR,
YFS3$\in$KORALZ (1990-1998), \Ordex{\alpha^1+h.o.LL} ISR+FSR, and
\KK\ MC (98-02) \Ordex{\alpha^2+h.o.LL} ISR+FSR+IFI with 
$d\sigma/\sigma = 0.2\%$; (2), for 
$e^+e^- \to e^+e^-+n\gamma$ for $\theta < 6^\circ$ there are 
BHLUMI 1.x, (1987-1990), \Ordex{\alpha^1} and
BHLUMI 2.x,4.x, (1990-1996), \Ordex{\alpha^1+h.o.LL} 
with $d\sigma/\sigma = 0.061\%$; (3), for 
$e^+e^- \to e^+e^-+n\gamma$ for $\theta > 6^\circ$ there
is BHWIDE (1994-1998), \Ordex{\alpha^1+h.o.LL} with
$d\sigma/\sigma = 0.2(0.5)\%$ at the Z peak ( just off the Z peak ); (4),
for $e^+e^- \to W^+W^-+n\gamma$, $W^\pm \to f\bar{f}$ there is
KORALW (1994-2001); and, (5), for 
$e^+e^- \to W^+W^-+n\gamma$, $W^\pm \to f\bar{f}$ there is 
YFSWW3 (1995-2001), YFS exponentiation + Leading Pole Approximation
with  $d\sigma/\sigma = 0.4\%$ at LEP2 energies above the WW threshold.
The typical MC realization we effect in Refs.~\cite{eex} is in the form of the
``matrix element $\times$ exact phase space'' principle,
as we illustrate in the following diagram:
\vspace{-4mm}
\begin{center}
{\includegraphics[height=50mm,width=50mm]{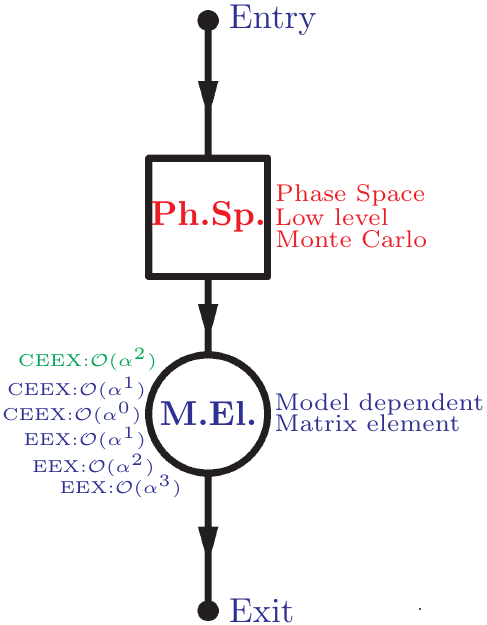}}
\end{center}

In practice it means the following:
\begin{itemize}
\item The universal exact Phase-space MC simulator is a separate module
      producing ``raw events'' (with importance sampling).
\item The library of several types of SM/QED matrix elements which provides
      the ``model weight'' is another independent module ( the \KK MC example is shown).
\item Tau decays and hadronization come afterwards of course.
\end{itemize}

The main steps in YFS exponentiation are the reorganization of the perturbative complete \Order{\alpha^\infty} series such that IR-finite $\bbeta$ components are isolated (factorization theorem) and 
the truncation of the IR-finite $\bbeta$s to finite \Order{\alpha^n}
with the attendant calculation of them from Feynman diagrams recursively.
We illustrate here the respective factorization for 
overlapping IR divergences for 
the 2$\gamma$ case -- $R_{12}\in R_1$ and $R_{12}\in R_2$ as they are
shown in the following picture:
\vspace{-4mm}
\begin{center}
{\includegraphics[width=45mm,angle=0]{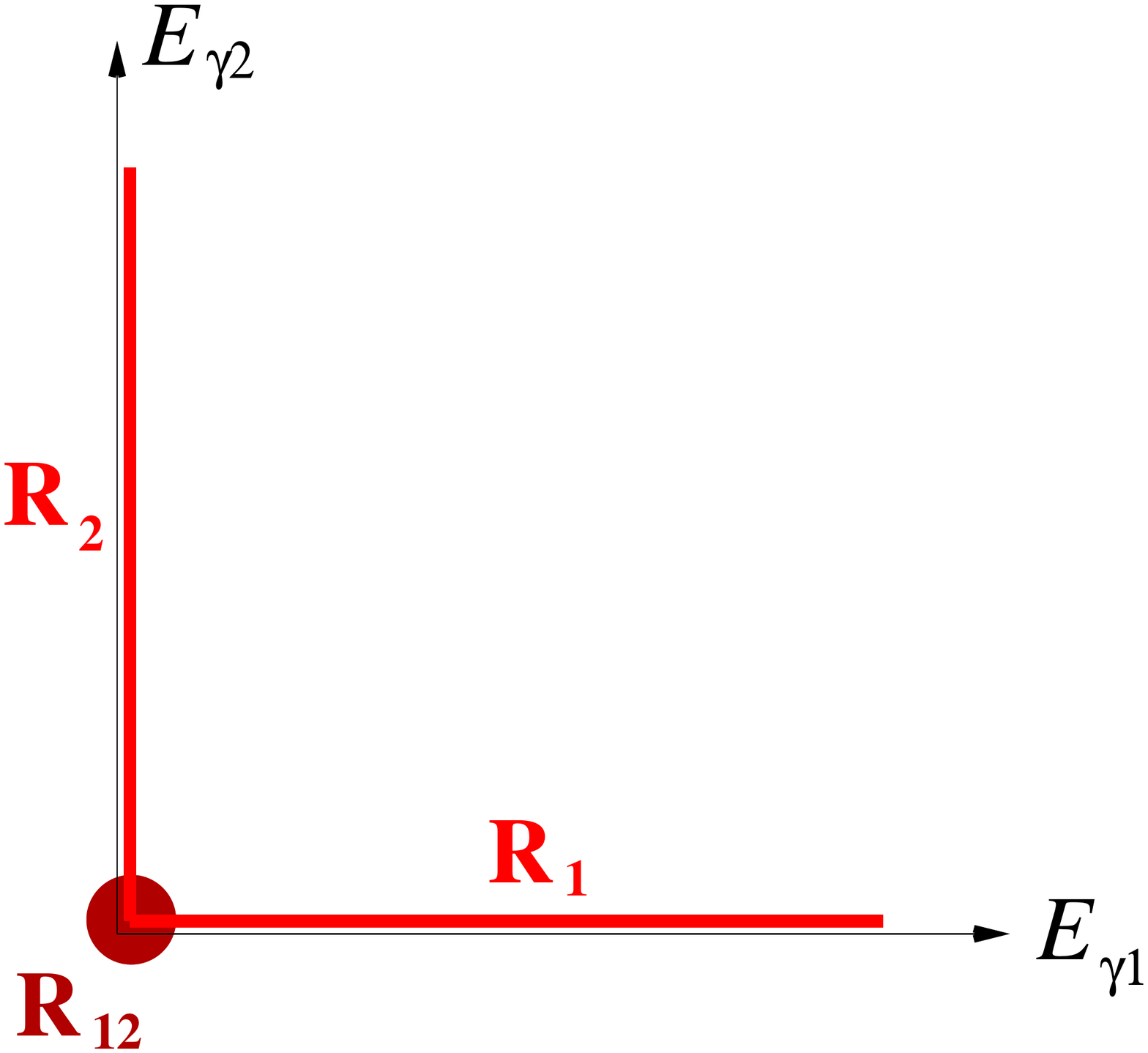}}
\end{center}
\noindent
$D_0(p_{f_1},p_{f_2},p_{f_3},p_{f_4}) =
      \bbeta_0(p_{f_1},p_{f_2},p_{f_3},p_{f_4})
;\\ ~~~~~~ p_{f_1}+p_{f_2}=p_{f_3}+p_{f_4}$\\ 
$D_1(p_f;k_1) = 
        \bbeta_0(p_f) \tilde{S}(k_1) +\bbeta_1(p_f;k_1) 
;\\ ~~~~~~ p_{f_1}+p_{f_2}\neq p_{f_3}+p_{f_4}$\\
$D_2(k_1,k_2) =
        \bbeta_0 \tilde{S}(k_1)\tilde{S}(k_2) 
       +\bbeta_1(k_1)\tilde{S}(k_2)+\bbeta_1(k_2)\tilde{S}(k_1)
       +\bbeta_2(k_1,k_2)$.\\
\noindent Note:
$\bbeta_0$ and $\bbeta_1$ are used beyond their usual (Born and $1\gamma$) 
respective phase spaces. A kind of smooth ``extrapolation'' or ``projection'' is always necessary. We see that a recursive order-by-order calculation of the 
IR-finite $\bbeta$s to a given fixed \Order{\alpha^n} is possible: specifically,\\
$\bbeta_0(p_{f_1},p_{f_2},p_{f_3},p_{f_4})= D_0(p_{f_1},p_{f_2},p_{f_3},p_{f_4})$,\\
$\bbeta_1(p_f;k_1) = 
        D_1(p_f;k_1) -\bbeta_0(p_f) \tilde{S}(k_1) $,\\
$\bbeta_2(k_1,k_2) = D_2(k_1,k_2)
       -\bbeta_0 \tilde{S}(k_1)\tilde{S}(k_2) 
       -\bbeta_1(k_1)\tilde{S}(k_2)-\bbeta_1(k_2)\tilde{S}(k_1)$,
$\ldots$,
allow such a truncation.

In the classic EEX/YFS schematically the $\beta$'s are truncated 
to \Order{\alpha^1}, in the ISR example. 
For
$e^-(p_1,\lambda_1)+e^+(p_2,\lambda_2)
 \to f(q_1,\lambda'_1)+\bar{f}(q_2,\lambda'_2)
 +\gamma(k_1,\sigma_1)+...+\gamma(k_n,\sigma_n)$,
we have
\begin{equation}
\sigma = \sum\limits_{n=0}^\infty \;\;
          \int\limits_{ m_\gamma} 
          d\Phi_{n+2}\; e^{Y(m_\gamma)} 
          D_n(q_1,q_2,k_1,...,k_n)
\label{eqeex1}
\end{equation}
with
\begin{equation}
\begin{split}
&D_0 = \bbeta_0,\qquad
D_1(k_1) = \bbeta_0 \tilde{S}(k_1) +\bbeta_1(k_1),
\\&
D_2(k_1,k_2) = \bbeta_0 \tilde{S}(k_1)\tilde{S}(k_2)
                    +\bbeta_1(k_1)\tilde{S}(k_2)
+\bbeta_1(k_2)\tilde{S}(k_1),
\\&
D_n(k_1,k_2...k_n)=\bbeta_0 \tilde{S}(k_1)\tilde{S}(k_2)...\tilde{S}(k_n)
\\&~
    +\bbeta_1(k_1)\tilde{S}(k_2)\tilde{S}(k_3)...\tilde{S}(k_n)
    +\tilde{S}(k_1)\bbeta_1(k_2)\tilde{S}(k_3)...\tilde{S}(k_n)
\\&~
 +...+\tilde{S}(k_1)\tilde{S}(k_2)\tilde{S}(k_3)...\bbeta_1(k_n).
\end{split}
\end{equation}
The real soft factors
and the IR-finite building blocks are
\begin{equation}
\begin{split}
&
\tilde{S}(k) = \sum\limits_\sigma |{\sfac}_\sigma(k)|^2 = |{\sfac}_+(k)|^2 +|{\sfac}_-(k)|^2
\\&~~~~~~~~~~~~~
             = -{\alpha\over 4\pi^2}\big({q_1\over kq_1}-{q_2\over kq_2} \big)^2
\\&
\bbeta_0= ( e^{-2\alpha\Re B_4} 
   \sum_\lambda |{\Meu}^{\rm Born+Virt.}_\lambda|^2 )\big|_{_{{\cal O}(\alpha^1)}},
\\&
\bbeta_1(k)=  \sum\limits_{\lambda\sigma} |{\Meu}^{\rm 1-PHOT}_{\lambda\sigma}|^2 
                  -\sum\limits_{\sigma}  |{\sfac}_\sigma(k)|^2
                   \sum\limits_{\lambda} |{\Meu}^{\rm Born}_\lambda|^2,
\end{split}
\end{equation}
with $\lambda$ = fermion helicity, $\sigma$ = photon helicity, 
and everything being in terms of $\sum_{spin} |...|^2 $!

The newer CEEX replaces older the EEX, where both are derived from 
the YFS theory~\cite{yfs}: EEX, Exclusive EXponentiation, 
is very close to the original Yennie-Frautschi-Suura formulation,
which is also now featured in the MC's Herwig++~\cite{hw++} and Sherpa~\cite{shpa} for particle decays. We need to stress that 
CEEX, Coherent EXclusive exponentiation, is an extension of the YFS
theory. Because of its coherence CEEX is friendly to quantum coherence
among the Feynman diagrams, so that we have the complete
$|\sum_{diagr.}^n {\Meu}_i\big|^2$ rather than the often incomplete
$\sum_{i,j}^{n^2} {\Meu}_i {{\Meu}_j}^*$. It follows that 
we get readily the proper treatment of
narrow resonances, $\gamma\oplus Z$~exchanges, $t\oplus s$~channels,
ISR$\oplus$FSR, angular ordering,~etc. KORALZ/YFS2, BHLUMI, BHWIDE, YFSWW, KoralW and  KORALZ are examples of the 
EEX formulation in our MC event generator approach; \KK MC is
the only example of the CEEX formulation.\par
 
Using the example of ISR \Order{\alpha^1} we illustrate CEEX schematically  
for the process
$e^-(p_1,\lambda_1)+e^+(p_2,\lambda_2) \to f(q_1,\lambda'_1)+\bar{f}(q_2,\lambda'_2)
+\gamma(k_1,\sigma_1)+...+\gamma(k_n,\sigma_n).$
We have
\begin{widetext}
\begin{equation}
\begin{split}
&\sigma = \sum\limits_{n=0}^\infty\;
 \int_{m_\gamma} d\Phi_{n+2}\!\!\!\!
 \sum\limits_{\lambda,\sigma_1,...,\sigma_n}\!\!\!\!
 |e^{\alpha B(m_\gamma)}{\Meu}^{\lambda}_{n,\sigma_1,...,\sigma_n}(k_1,...,k_n)|^2,
\qquad
{\Meu}_{0}^{\lambda} = \hbeta_0^\lambda,\quad
{\Meu}^\lambda_{1,\sigma_1}(k_1) 
       = \hbeta^\lambda_0 {\sfac}_{\sigma_1}(k_1) 
         +\hbeta^\lambda_{1,\sigma_1}(k_1),\qquad
\\&
{\Meu}^\lambda_{2,\sigma_1,\sigma_2}(k_1,k_2) 
= \hbeta^\lambda_0 {\sfac}_{\sigma_1}(k_1) {\sfac}_{\sigma_2}(k_2)
   +\hbeta^\lambda_{1,\sigma_1}(k_1){\sfac}_{\sigma_2}(k_2)
   +\hbeta^\lambda_{1,\sigma_2}(k_2){\sfac}_{\sigma_1}(k_1),
\qquad
{\Meu}^\lambda_{n,\sigma_1,...\sigma_n}(k_1,...k_n) 
= \hbeta^\lambda_0 {\sfac}_{\sigma_1}(k_1) {\sfac}_{\sigma_2}(k_2) ... {\sfac}_{\sigma_n}(k_n)+
\\&
  +\hbeta^\lambda_{1,\sigma_1}(k_1) {\sfac}_{\sigma_2}(k_2)...{\sfac}_{\sigma_n}(k_n)
  +{\sfac}_{\sigma_1}(k_1) 
     \hbeta^\lambda_{1,\sigma_2}(k_2)... {\sfac}_{\sigma_n}(k_n)+...
   + {\sfac}_{\sigma_1}(k_1) {\sfac}_{\sigma_2}(k_2)...
      {\sfac}_{\sigma_{n-1}}(k_{n-1}) \hbeta^\lambda_{1,\sigma_n}(k_n),
\end{split}
\end{equation}
\end{widetext}
where $\lambda$ is the collective index of fermion helicities.
The \Order{\alpha^1} IR-finite building blocks are:
\[
\begin{split}
&\hbeta^\lambda_0 = \big(e^{-\alpha B_4} 
  {\Meu}^{\rm Born+Virt.}_{\lambda}\big)\big|_{{\cal O}(\alpha^1)},
\\&
\hbeta^{\lambda}_{1,\sigma}(k)=
  {\Meu}^\lambda_{1,\sigma}(k) - \hbeta^\lambda_0 {\sfac}_{\sigma}(k)
\end{split}
\]
Everything above is expressed in terms of ${\Meu}$-amplitudes!
Distributions are $\geq 0$ by construction!
In \KK MC the above is done up to \Order{\alpha^2} for ISR and FSR.

The full scale CEEX \Order{\alpha^r}, r=1,2, master formula 
for the polarized total cross section reads as follows:
\begin{equation}
\label{eqceex1}
\begin{split}
&\sigma^{(r)}\! = \!\!
  \sum_{n=0}^\infty {1\over n!}
  \int d\tau_{n} ( p_a\!+\!p_b ; p_c,p_d, k_1,\dots,k_n)
\\&~\times
 e^{ 2\alpha\Re B_4 }\!
    \sum_{\sigma_i,\lambda,\bar{\lambda}}\;
    \sum_{i,j,l,m=0}^3
    \hat{\varepsilon}^i_a \hat{\varepsilon}^j_b
    \sigma^i_{\lambda_a \bar{\lambda}_a} 
    \sigma^j_{\lambda_b \bar{\lambda}_b}
\\&~\times
 \Mmf^{(r)}_n 
    \left(\st^{p}_{\lambda} \st^{k_1}_{\sigma_1} \st^{k_2}_{\sigma_2}
                                           \dots \st^{k_n}_{\sigma_n} \right)
    \Big[\Mmf^{(r)}_n\!
     \left(  \st^{p}_{\bar{\lambda}} \st^{k_1}_{\sigma_1} 
             \st^{k_2}_{\sigma_2}\dots \st^{k_n}_{\sigma_n}
     \right)\Big]^\star
 \sigma^l_{\bar{\lambda}_c \lambda_c } 
   \sigma^m_{\bar{\lambda}_d \lambda_d } 
   \hat{h}^l_c              \hat{h}^m_d.
\end{split}
\end{equation}

The respective CEEX amplitudes are
\begin{widetext}
\begin{equation}
\begin{split}
\Mmf^{(1)}_n\left(\st^{p}_{\lambda} \st^{k_1}_{\sigma_1}
                 \dots \st^{k_n}_{\sigma_n}\right)\!
  &=\sum\limits_{\wp\in{\cal P}}
   \prod\limits_{i=1}^n  {\sfac}_{[i]}^{\{\wp_i\}}
   \Bigg\{ \hat\beta_0^{(1)}\big(\st^{p}_{\lambda};X_\wp\big)\! 
+\sum\limits_{j=1}^n
    {\hat\beta^{(1)}_{1\{\wp_j\}}
    \big(\st^{p}_{\lambda}\st^{k_j}_{\sigma_j};X_{\wp}\big) 
             \over {\sfac}_{[j]}^{\{\wp_j\}} }
   \Bigg\}
\\
\Mmf^{(2)}_n\left(\st^{p}_{\lambda} \st^{k_1}_{\sigma_1}
                              \dots \st^{k_n}_{\sigma_n}\right)
&=\!\sum\limits_{\wp\in{\cal P}}
   \prod\limits_{i=1}^n  {\sfac}_{[i]}^{\{\wp_i\}}
   \Bigg\{ \hat\beta_0^{(2)}\big(\st^{p}_{\lambda};X_\wp\big) 
   \!+\! { \sum\limits_{j=1}^n}
   {\hat\beta^{(2)}_{1\{\wp_j\}}\big(\st^{p}_{\lambda}\st^{k_j}_{\sigma_j};X_{\wp}\big) 
             \over {\sfac}_{[j]}^{\{\wp_j\}} }\!
 +\! { \sum\limits_{1\leq j<l\leq n}}
   {\hat\beta^{(2)}_{2\{\wp_j,\wp_l\}}
    \big( \st^{p}_{\lambda}\st^{k_j}_{\sigma_j} \st^{k_l}_{\sigma_l}; X_{\wp}\big) 
         \over {\sfac}_{[j]}^{\{\wp_j\}}  {\sfac}_{[l]}^{\{\wp_l\}} }
   \Bigg\}.
\end{split}
\end{equation}
\end{widetext}
For the full details see ref.~\cite{ceex1}.

\begin{figure}[h]
\centering
{\includegraphics[width=70mm]{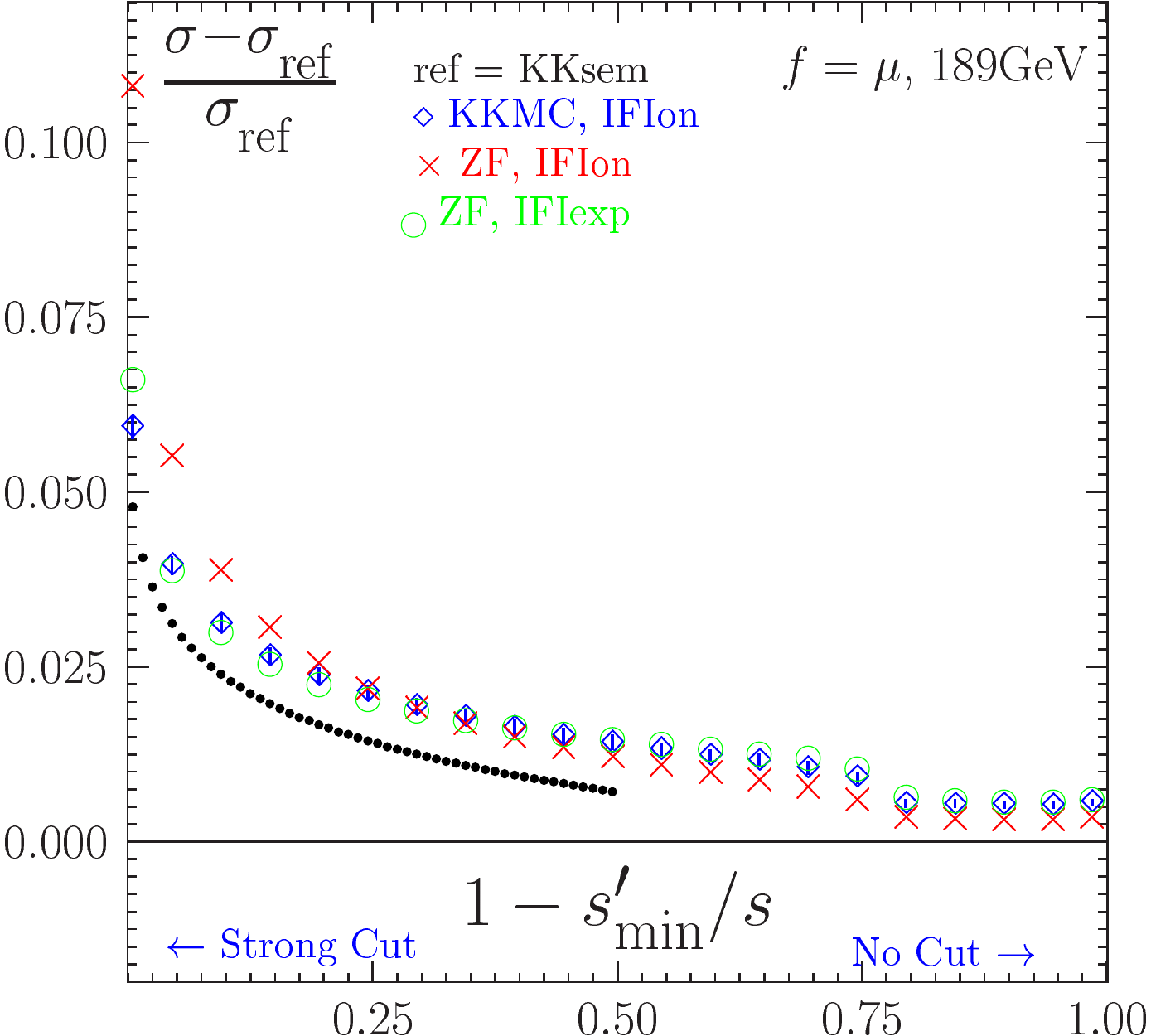}}
\caption{Principal cross checks of \KK MC for
$e^-e^+\to \mu^-\mu^+ +n\gamma$ process at $\sqrt{s}=$189GeV.}
\label{zfit}
\end{figure}
The precision tags of the \KK MC are determined by comparisons with
our own semi-analytical and independent MC results and by comparison
with the semi-analytical results of the program ZFITTER~\cite{zfitter}.
In Fig.~\ref{zfit} we illustrate such comparisons, which lead to
the \KK MC precision tag $d\sigma/\sigma = 0.2\%$ for example.
The ISR of ZFITTER is based on the \Order{\alpha^2}
result of ref.~\cite{berends},
while \KK MC is totally independent!
See Ref.~\cite{ceex1,recent2} for a more complete discussion.
Thus, we know that \KK MC has the capability to deliver per mille precision
on the large EW effects if it is extended to the appropriate 
incoming beams for the LHC and the muon collider. To this we now turn.\par

\section{Extension of \KK MC to the processes
 $f\bar{f}\to f'\bar{f'},$ 
 $f=\mu,q,\nu_\ell$, 
 $f'=\ell, \nu_\ell, q , q=u,d,s,c,b,$
 $\ell=e,\mu,\tau, f\ne f', $}

At the LHC and at a futuristic muon collider~\cite{muclldr}, 
the incoming beams involve for $Z/\gamma*$ production 
and decay the other light charged fundamental fermions in the SM: $u,d,s,c,b$ for the LHC and the muon for a muon collider.
Thus, we need to extend the matrix elements, residuals, and IR functions in (\ref{eqeex1},\ref{eqceex1}) to the case where we substitute the $e^-,\; e^+$ EW charges by the new beam particles
$f,\; \bar{f}$ EW charges and we substitute the mass $m_e$ everywhere by $m_f$
\footnote{We advise the reader that especially in the QED radiation module KarLud 
for the ISR in \KK MC, see Ref.~\cite{kkmc}, some of the expressions had
$Q_e$ and $m_e$ effectively hard-wired into them and these had to all be found
and substituted properly.}. 
We have done this with considerable cross checks against the same semi-analytical tools that we employed in Ref.~\cite{ceex1} to establish the precision tag of version 4.13 of \KK MC. We want to stress that this was a highly non-trivial set of cross-checks: for example, we found that the MC procedure used in the crude MC cross section was unstable when the value of the radiation strength factor 
$\gamma_f=\frac{2Q_F^2\alpha}{\pi}\left(\ln(s/m_f^2)-1\right)$
becomes too small\footnote{In the case of the quarks, we will use here the current quark mass values $m_u\cong 5$MeV and $m_d\cong 10$MeV following Ref.~\cite{mrst} for our illustrations; we leave these
values as user input in general.}. This instability was removed and the correct value of the MC crude cross section was verified by semi-analytical methods. We did therefore a series of cross checks/illustrations with the new version of \KK MC, version 4.22, which we now exhibit.\par
\begin{figure}[h]
\centering
\includegraphics[width=70mm]{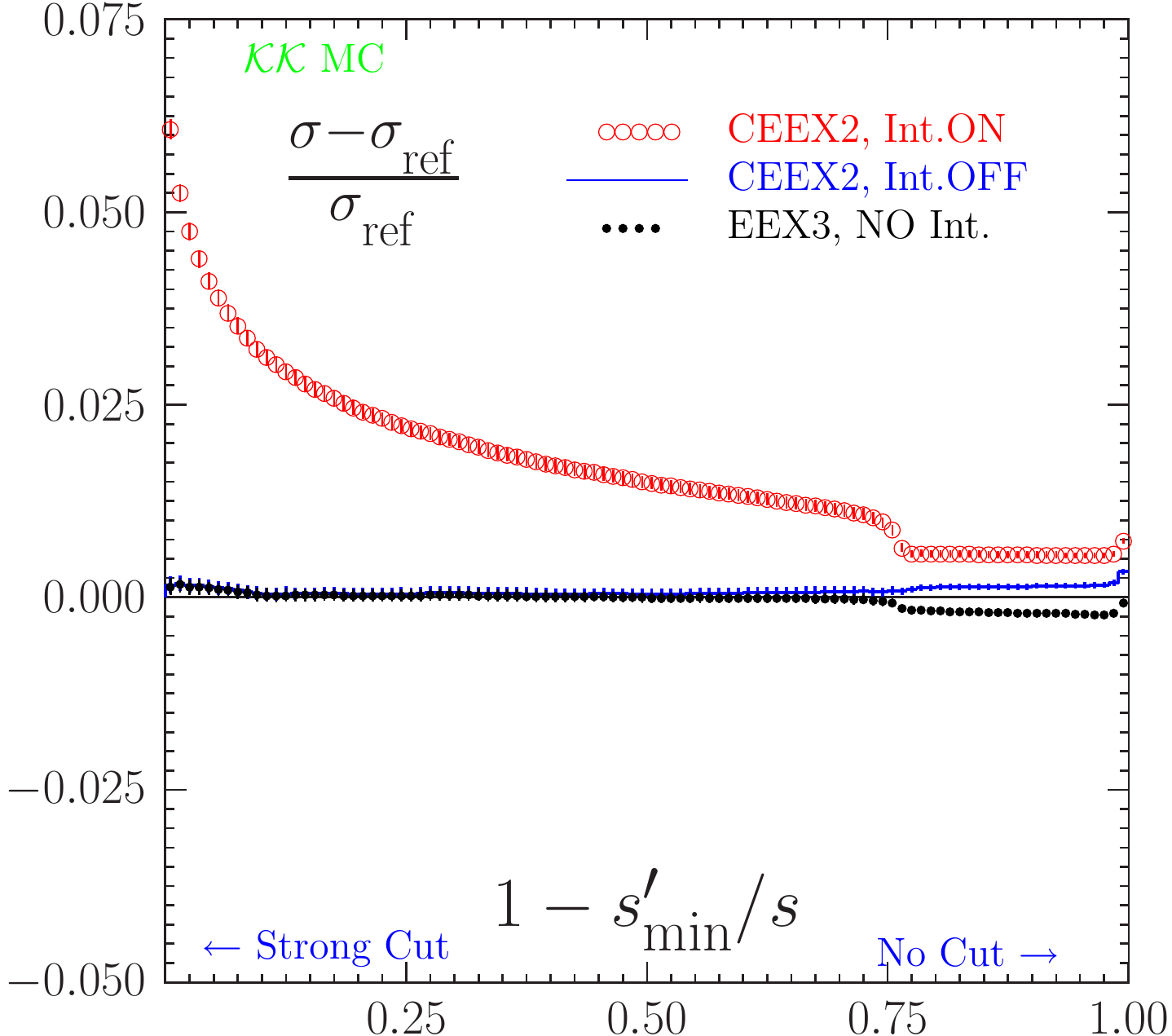}
\caption{
Total cross section $\sigma$, energy cut-off study.
The same results as in the Table~\ref{tab:table1}.
Ref. $\sigma_{\rm ref}$ = semianalytical of \KK{}sem.
}
\label{fig2}
\end{figure}
\begin{figure}[h]
\centering
\includegraphics[width=70mm]{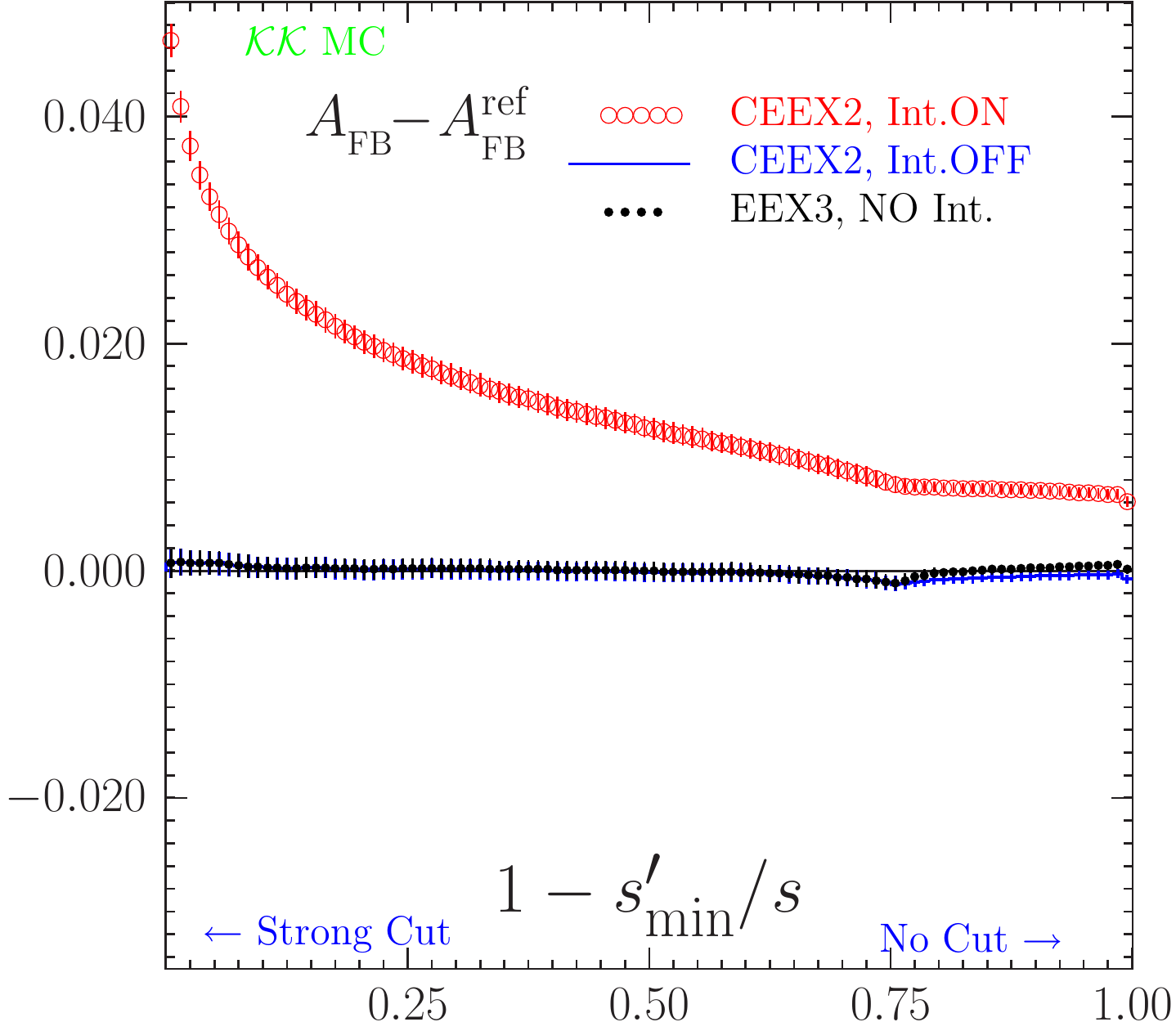}
\caption{
Energy cut-off study of charge asymmetry $A_{\rm FB}$ for
the process $e^+e^-\to\mu^+\mu^-$.
The same results as in the Table~\ref{tab:table1}.
Reference $A_{\rm FB}^{\rm ref}$ =  semianalytical \KK{}sem.
}
\label{fig3}
\end{figure}

Turning first to the most important cross-check, we show in Tab.~\ref{tab:table1} and Figs.~\ref{fig2}-\ref{fig4} that for the $e^+e^-\rightarrow \mu^+\mu^-$ process, our new version \KK MC 4.22 reproduces the results in the corresponding $\sqrt{s}=189$GeV
cross checks done in Ref.~\cite{ceex1} for the dependence of the CEEX calculated cross section and $A_{FB}$ on the energy cut-off on $v=1-s'/s$ where $s'=M^2_{\mu\bar{\mu}}$ is the invariant mass of the $\mu\bar{\mu}$-system. 
The reader can check that the two sets of results, those given here in Tab.~\ref{tab:table1} and Figs.~\ref{fig2}-\ref{fig4} and those given in Table 5, Figs.~20,21, and 18 in Ref.~\cite{ceex1} are in complete agreement within statistical fluctuations. This shows that our introduction of the new beams has not spoiled the precision of the \KK MC for the incoming $e^+e^-$ state.

\def\Energy{189GeV}
\def\Process{$e^-e^+ \to \mu^-\mu^+$}
\def\Angle{$\theta^{\bullet}$}
\begin{table}[h]
\centering
{\includegraphics[width=90mm,height=60mm]{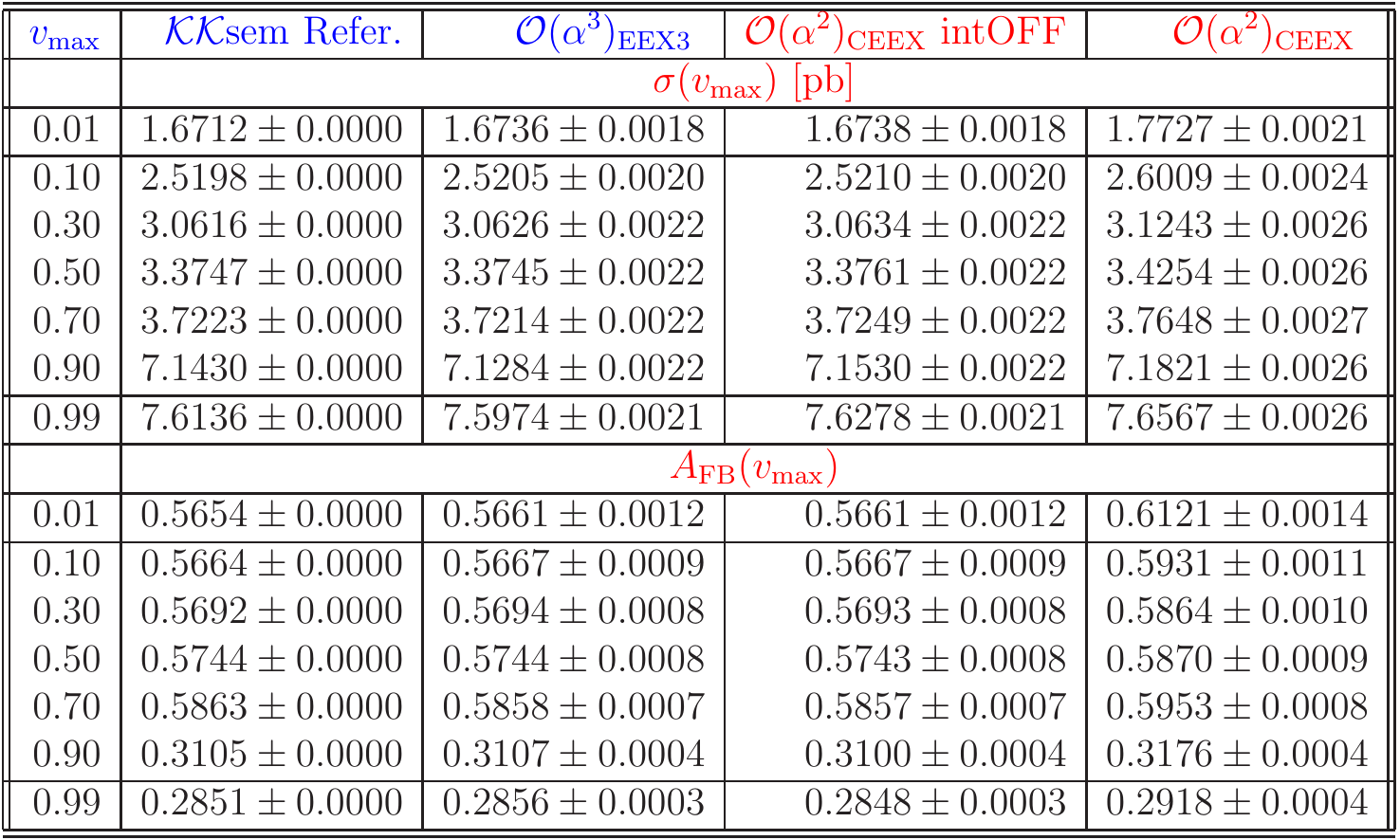}}
\caption{
 Energy cut-off study of
 total cross section  $\sigma$ and charge asymmetry $A_{\rm FB}$
 for annihilation process $e^-e^+ \to \mu^-\mu^+$, at $\sqrt{s}=$189GeV.
 Energy cut: $v<v_{\max}$, $v=1-M^2_{f\bar{f}}/s$.
 Scattering angle for $A_{\rm FB}$ is \Angle
 (defined in Phys. Rev. {\bf D41}, 1425 (1990)).
 No cut in \Angle.
 E-W corr. in \KK\  according to DIZET 6.x.
 In addition to CEEX matrix element, results are also shown for
 \OrderLL{\alpha^3} EEX3 matrix element without ISR$\otimes$FSR interf.
 \KK{}sem is semianalytical program, part of \KK{}MC.
}
\label{tab:table1}
\end{table}

\begin{figure}
\centering
\includegraphics[width=71mm,height=40mm]{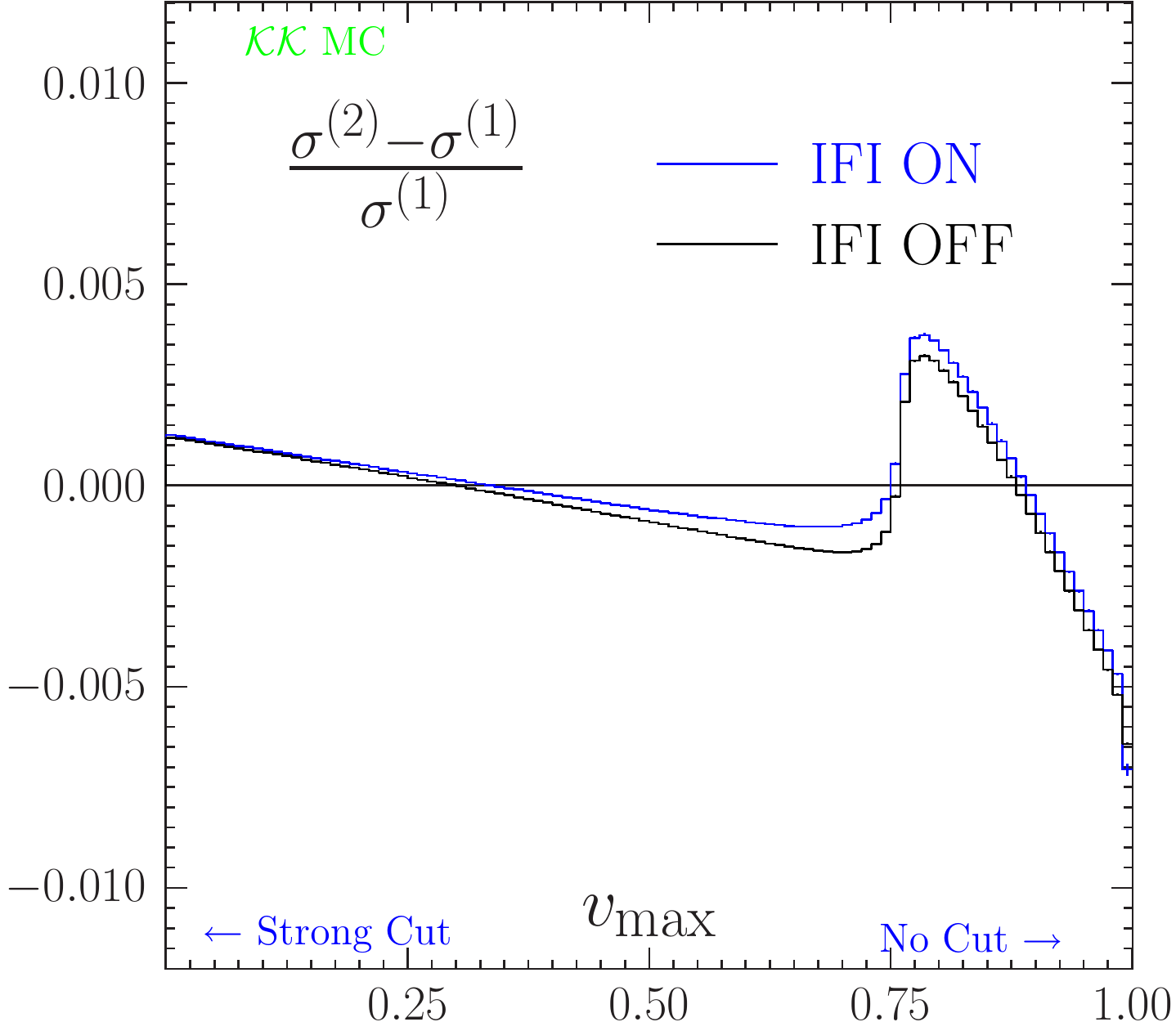}
\includegraphics[width=71mm,height=40mm]{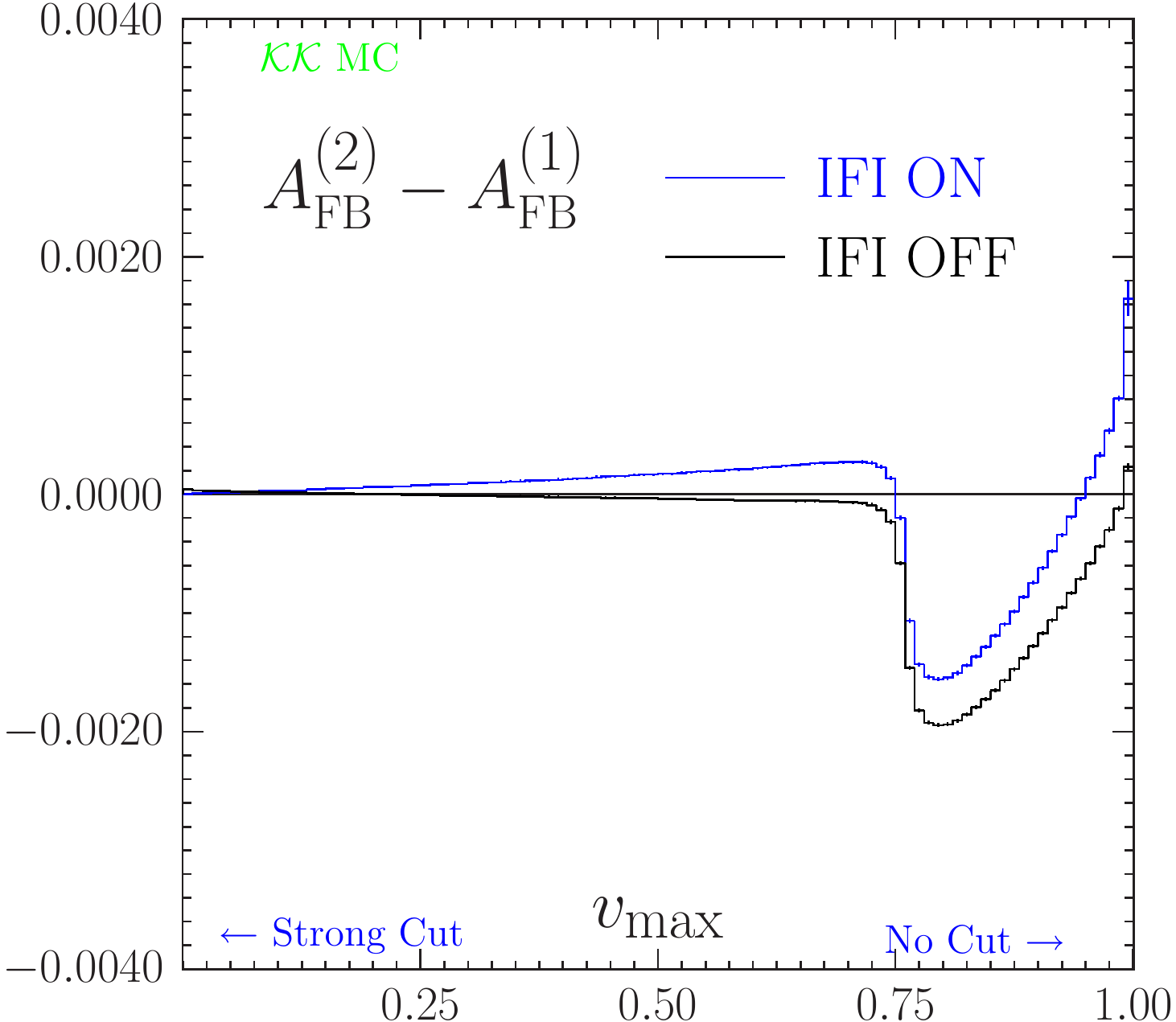}
\caption{
Physical precision of CEEX ISR matrix element
for \Process at $\sqrt{s}=$\Energy.
See table \ref{tab:table1} for definition
of cut-offs.
}
\label{fig4}
\end{figure}

We turn next to the new type of incoming beam scenario in Tab.~\ref{tab:table2} and Figs.~\ref{fig6}-\ref{fig8} wherein we show the analogous results to those in Tab.~\ref{tab:table1} and Figs.~\ref{fig2}-\ref{fig4}
for the process $d\bar{d}\rightarrow \mu^-\mu^+$ at $\sqrt{s}=189$GeV so that we can keep a good reference to the relative size of the EW corrections versus
what one would have in the usual $e^+e^-$ annihilation case.
We see that for strong cuts, with $v_{max}\sim .01$ and for the loose cut,
with $v_{max}\sim 0.99$, the effects are similar to those in the more
familiar incoming $e^+e^-$ annihilation case, as the sign of the EW charges are the same for the $d$ and the $e^-$. The values are different so that size of the effects in Tab.~\ref{tab:table2} and Figs.~\ref{fig6}-\ref{fig8} are correspondingly different. For example, in the strong cut, turning the initial-final state interference(IFI) off changes the CEEX cross section result for $v_{max}=0.01$ by $-1.9\%$ for the incoming $d\bar{d}$ case compared to $-5.9\%$ for the incoming $e^-e^+$ case. The behavior of $A_{FB}(v_{max})$
is similar between to the two incoming beam sets, where turning the IFI off reduces the value of $A_{FB}$ at $v_{max}=0.01$ by $8.12\%(2.55\%)$
respectively for the incoming $e^-e^+(d\bar{d})$ case. In both cases, the loose cut such as $v_{max}=0.99$ tends to wash-out these effects. In Fig.~\ref{fig6}
the data on the cross sections in the table in Tab.~\ref{tab:table2} are plotted in relation to the reference semi-analytical result denoted as \KK{sem}~\cite{ceex1}
as the ratio of their difference to the reference divided by the reference and in Fig.\ref{fig7} the corresponding data on $A_{FB}$ are plotted as their difference with the respective \KK{sem} results. When compared to the analogous results for the usual $e^-e^+$ case in Figs.~\ref{fig2} and \ref{fig3} we see that structure at the Z-radiative return position, $v_{mas}\cong 0.77$, is very much reduced in the $d\bar{d}$ case due to the smaller electric charge magnitude, just as the size of the IFI effects themselves are similarly reduced. In Fig.~\ref{fig8}, we show the physical precision test which compares the size of the second and first order CEEX results for the cross section and the forward-backward asymmetry:
for the $d\bar{d}$ case compared to the similar plots in Fig.~\ref{fig4} for the   $e^-e^+$ case we see that for the strong cuts we have higher precision, we have smooth behavior through the Z-peak region, and that at the very loose cuts the two precision tags are similar, where we would estimate that similar value at $0.35\%$ in the worst case that $v_{max} \rightarrow 1$ on the cross section for example -- here we use 
half the difference shown in the figure as the error estimate. For the more generic energy cut of $0.6\%$ our physical precision estimate is $0.05\%$. 
This is the type
of precision required for the precision LHC physics studies.

\def\Process{$d \bar{d} \to \mu^-\mu^+$}
\begin{table}[h]
\centering
\includegraphics[width=90mm,height=60mm]{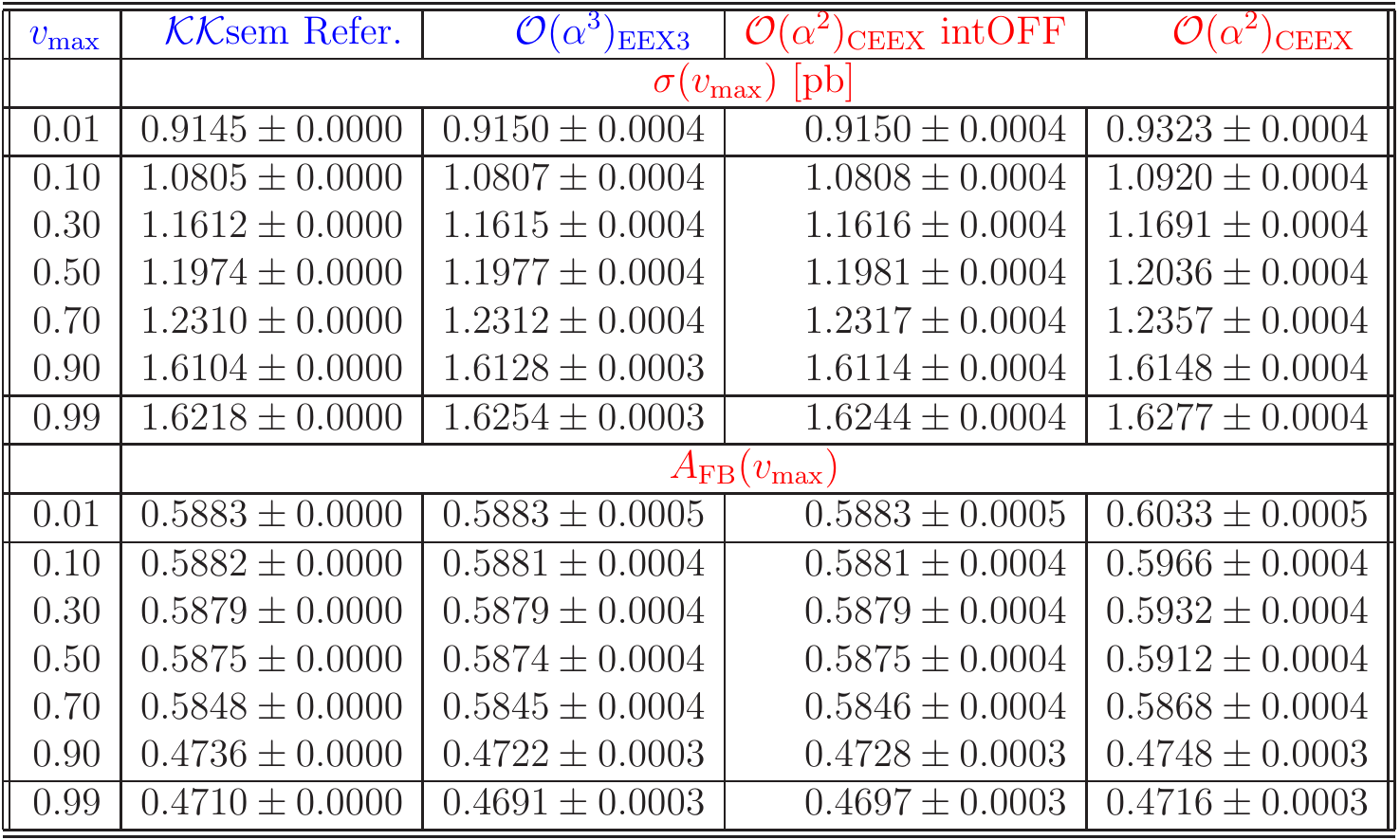}
\caption{
 Study of total cross section $\sigma(v_{\max})$ 
 and charge asymmetry $A_{\rm FB}(v_{\max})$,
 \Process, at $\sqrt{s}$~=\Energy.
 See Table \ref{tab:table1} for definition of
 the energy cut $v_{\max}$, scattering angle and M.E. type, 
}
\label{tab:table2}
\end{table}

\begin{figure}[h]
\centering
\includegraphics[width=70mm]{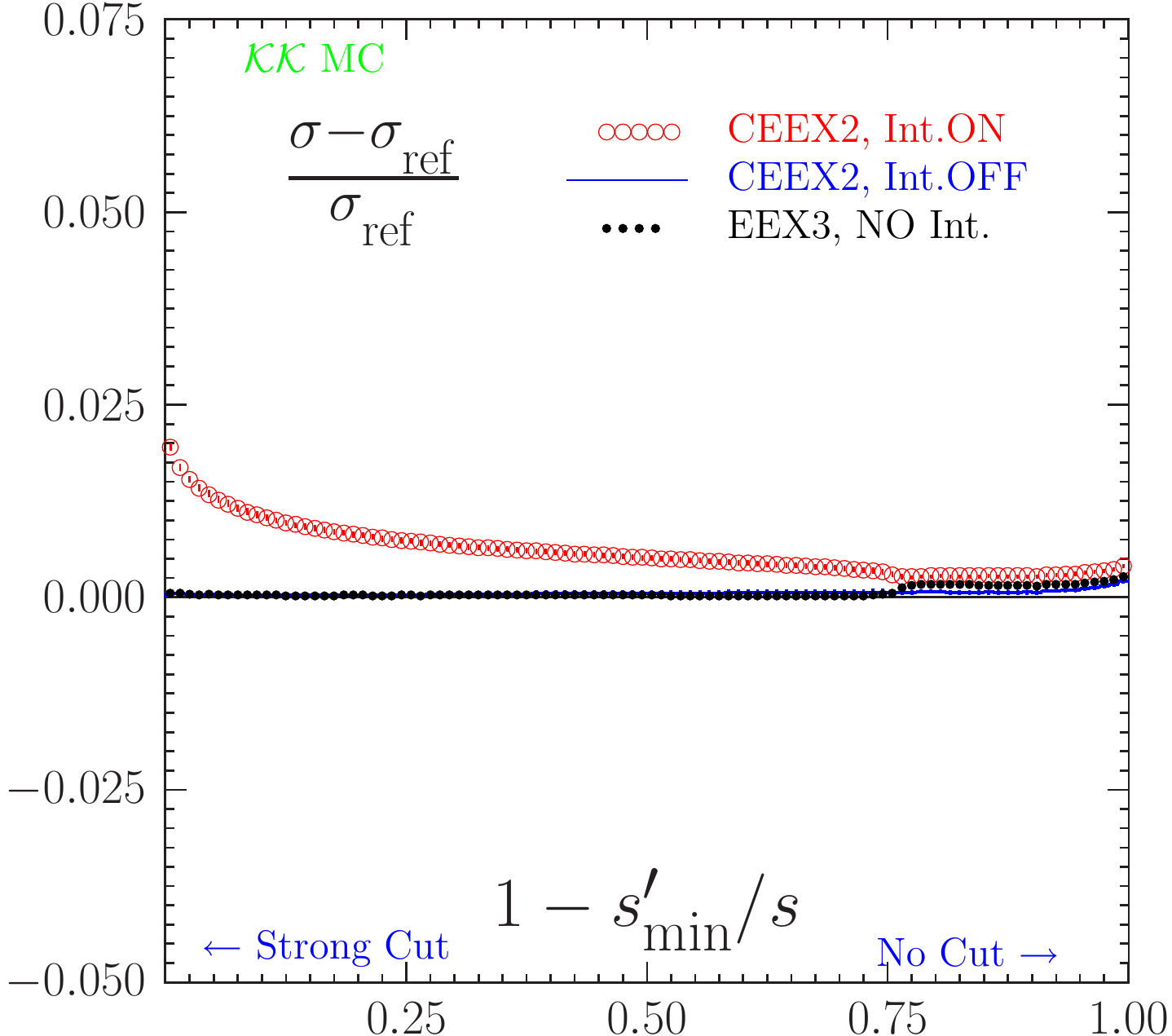}
\caption{
Energy cut-off study
of Total cross section for \Process, at \Energy.
The same as in the table \ref{tab:table2}.
$\sigma_{\rm ref}$ = semianalytical of \KK{}sem.
}
\label{fig6}
\end{figure}

\begin{figure}[h]
\centering
\includegraphics[width=70mm]{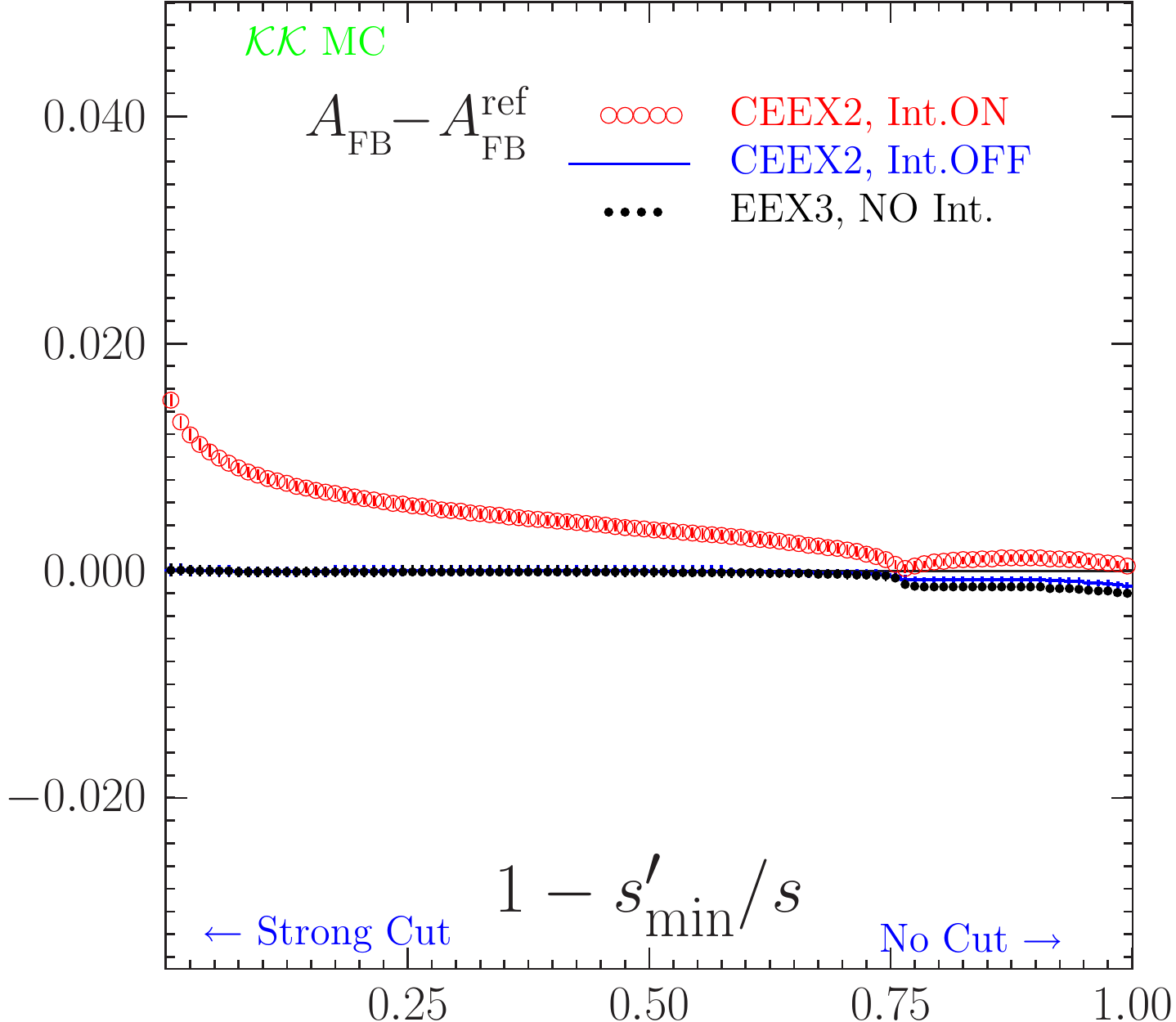}
\caption{
Energy cut-off study of charge asymmetry $A_{\rm FB}$ for
the process \Process, at \Energy.
Reference $A_{\rm FB}^{\rm ref}$ from semianalytical \KK{}sem.
}
\label{fig7}
\end{figure}

\begin{figure}
\centering
\includegraphics[width=71mm,height=40mm]{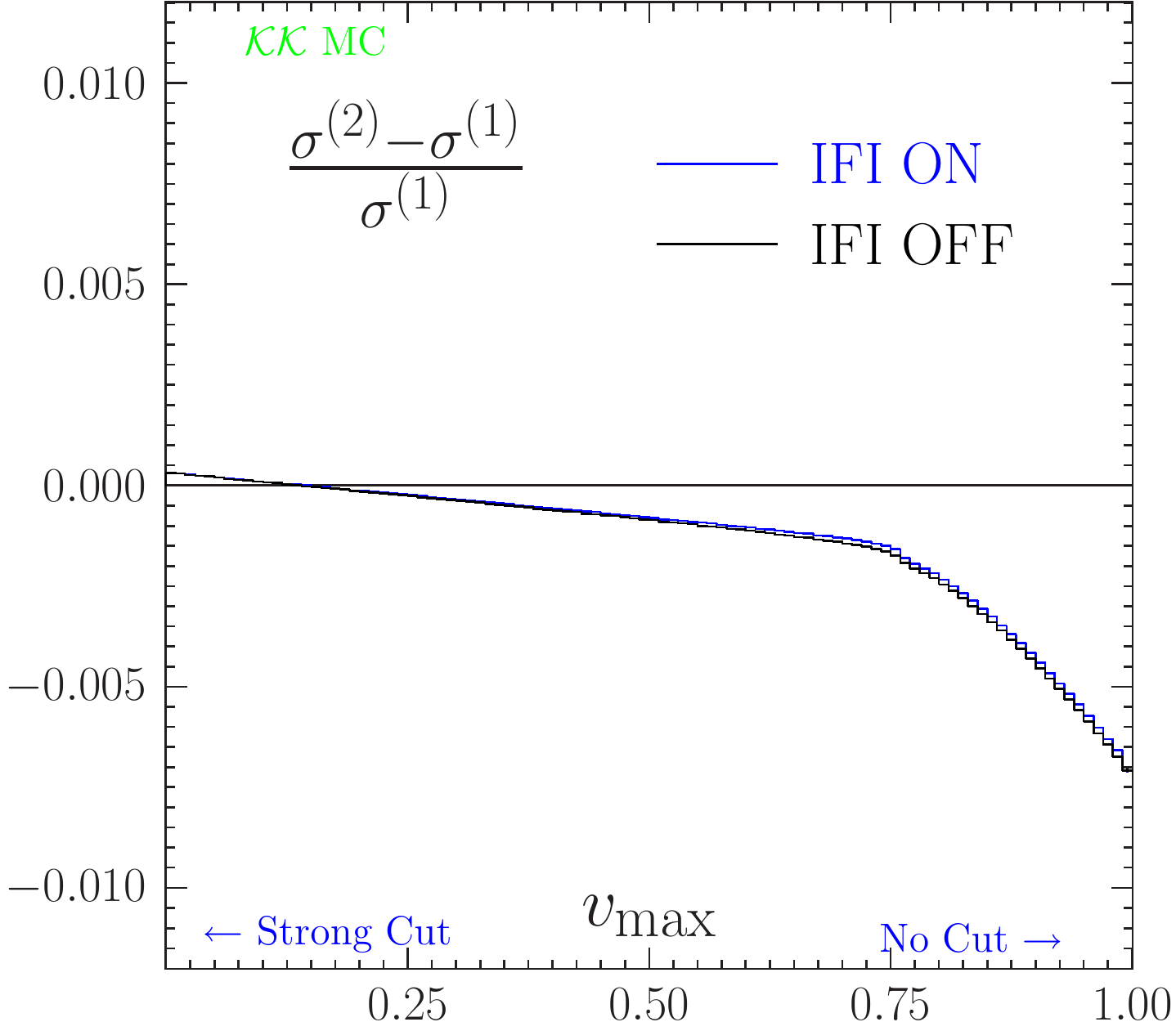}
\includegraphics[width=71mm,height=40mm]{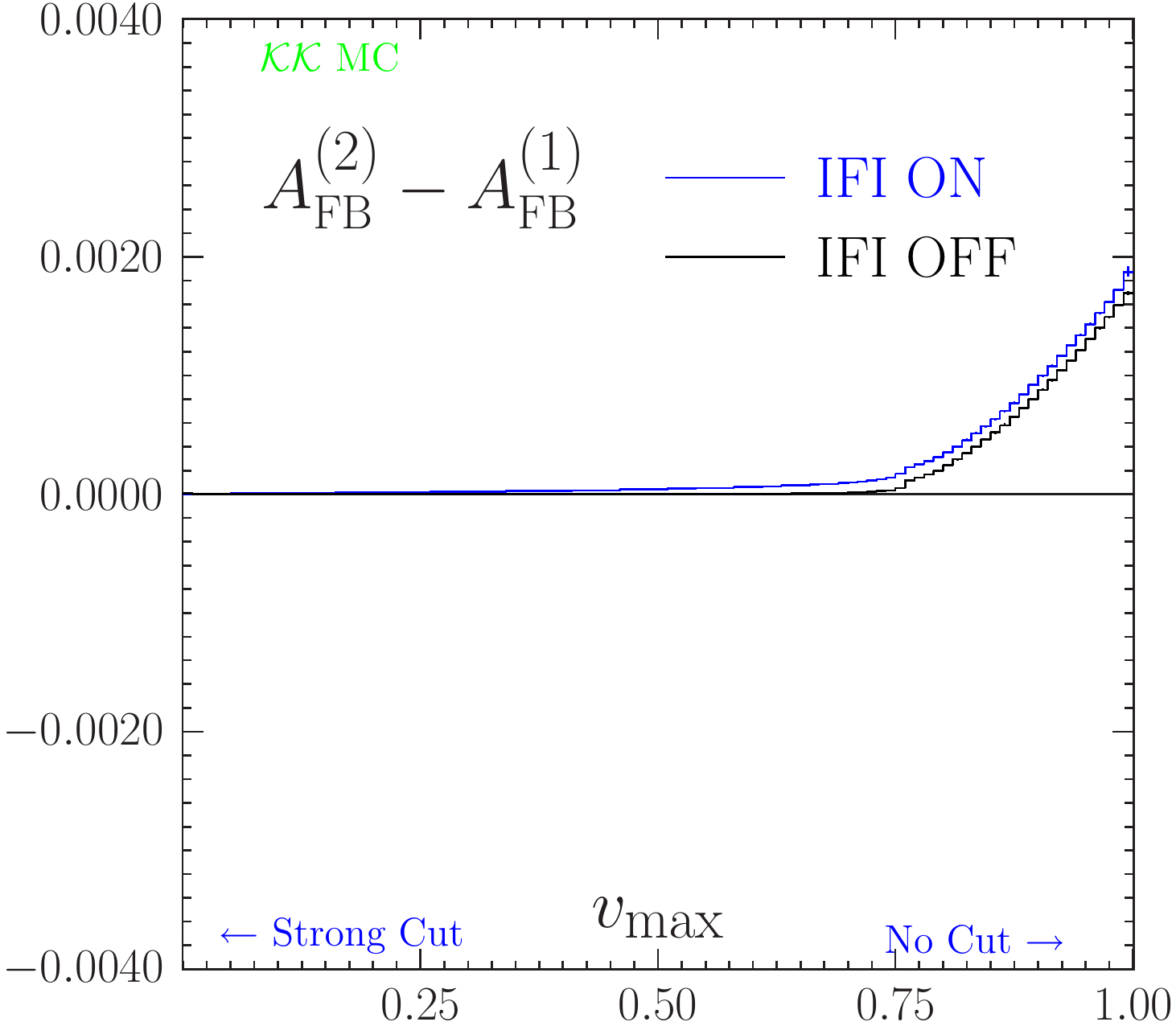}
\caption{
Physical precision of CEEX ISR matrix element
for \Process at $\sqrt{s}=$\Energy.
See table \ref{tab:table1} for definition
of cut-offs.
}
\label{fig8}
\end{figure}

\begin{figure}[h]
\centering
\includegraphics[width=70mm]{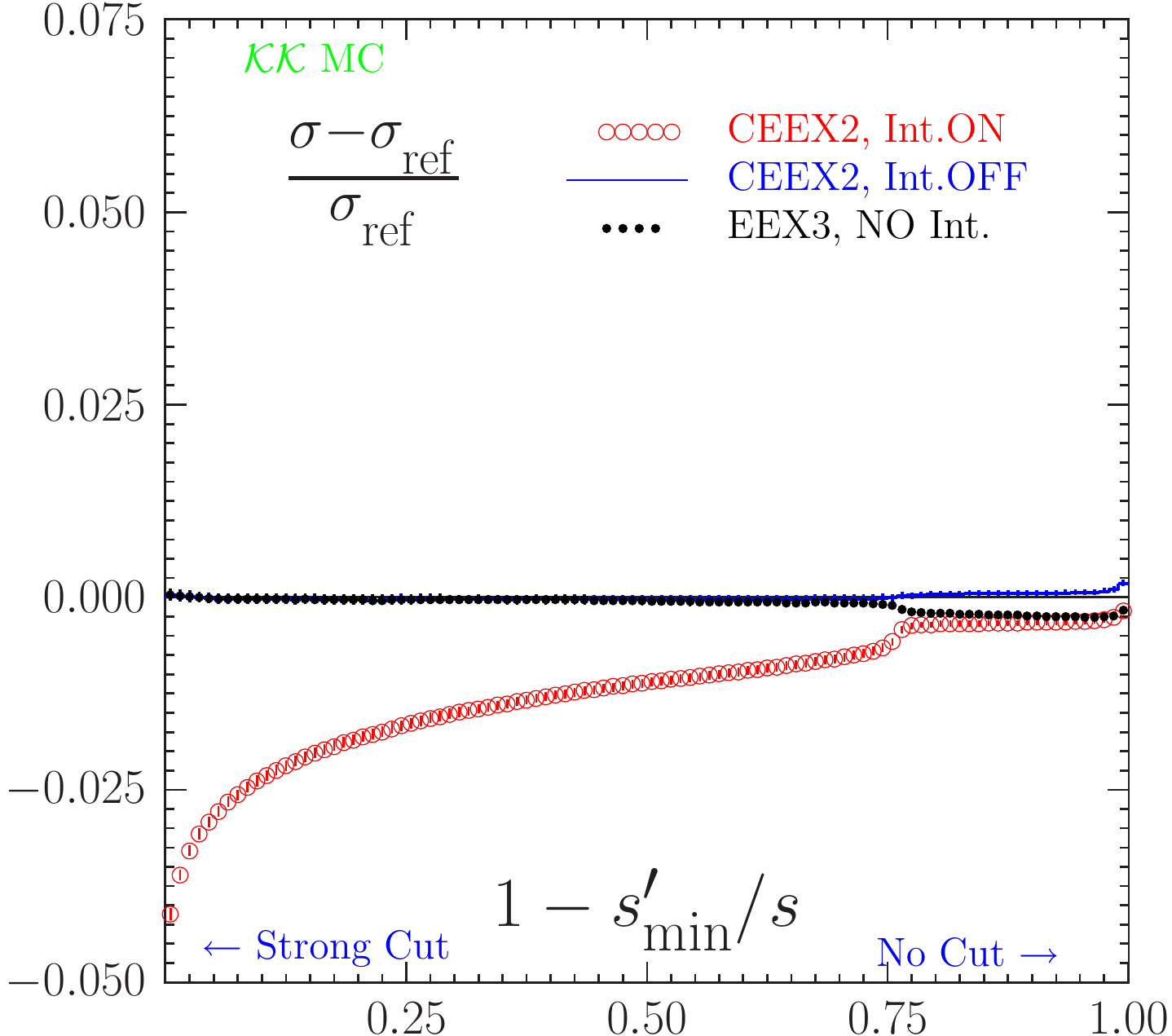}
\caption{
Total cross section $\sigma$, energy cut-off study for
the process $u\bar{u}\to\mu^+\mu^-$.
The same as in the table \ref{tab:table3}. No cut in \Angle.
Ref. $\sigma_{\rm ref}$ = semianalytical of \KK{}sem.
}
\label{fig10}
\end{figure}

\begin{figure}[h]
\centering
\includegraphics[width=70mm]{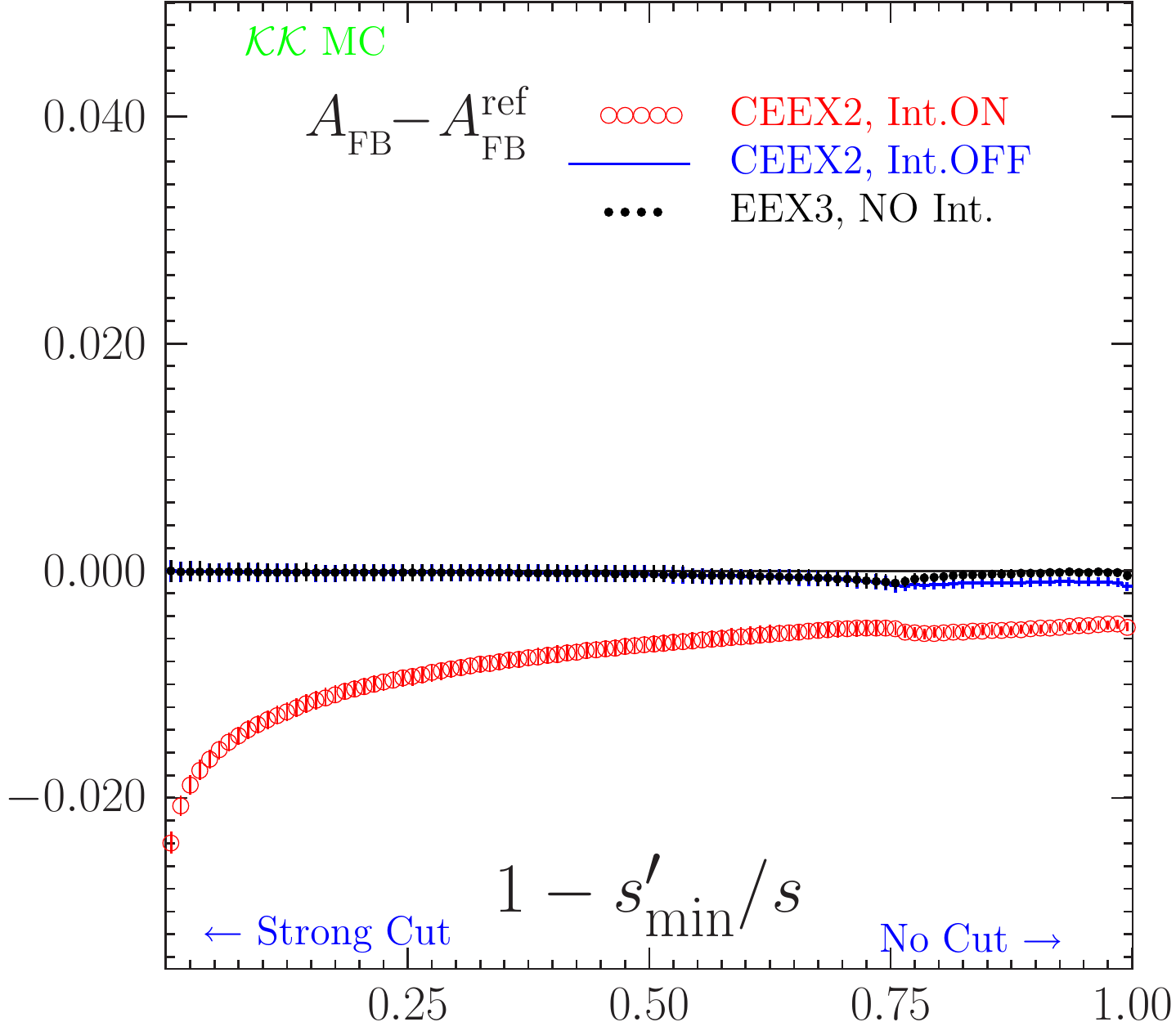}
\caption{
Charge asymmetry $A_{\rm FB}$, energy cut-off study for
the process $u\bar{u}\to\mu^+\mu^-$.
The same as in the table \ref{tab:table3}. No cut in \Angle.
}
\label{fig11}
\end{figure}

Turning next to the incoming $u\bar{u}$ case, we show in 
Tab.~\ref{tab:table3} and Figs.~\ref{fig10}-\ref{fig12} the analogous results to those in 
Tab.~\ref{tab:table2} and Figs.~\ref{fig6}-\ref{fig8} for the
 $u\bar{u}\rightarrow \mu^-\mu^+$ at $\sqrt{s}=189$GeV, so that again 
we have the reference to the usual
incoming $e^+e^-$ annihilation case regarding the size and nature of the 
EW effects expected. We see that the effects are now quantitatively different, because the sizes of the EW charges are different, but they also have the opposite sign in the enhanced regions because the EW charges of the u quarks have the opposite sign to those of the $e^-$. This means that in the LHC environment
in processes such as single $Z$ boson production there will be some 
compensation between the effects from u and d quarks. A detailed application of the new \KK MC two such scenarios will appear elsewhere. Here, we specifically note that for the strong cut case with $v_{max}=0.01$ the IFI effect on the cross section in Tab.~\ref{tab:table3} is $-4.14\%$  while the effect on $A_{FB}$ at
this value of $v_{max}$ is $-3.52\%$, both of which correlate well with the
value of the u-quark EW charges compared to the $e^-$ EW charges, where the corresponding results are from Tab.~\ref{tab:table1} $5.9\%$ and $8.12\%$ respectively.
In Figs.~\ref{fig10} and \ref{fig11} we show for the incoming $u\bar{u}$ the analogous plots to those in Figs.~\ref{fig6} and \ref{fig7} for the incoming
$d\bar{d}$ case of the relative values of the data in Tab.~\ref{tab:table3}. We see that the structure at the Z-radiative return position is a bit more evident than for the latter case and that the IFI(Initial-Final state Interference) effects are correspondingly more evident in general, as expected. In Fig.~\ref{fig12}, we show the corresponding physical precision study as the difference between the second and first order CEEX predictions.
In the worst case scenario with $v_{max}\rightarrow 1$ we have the estimate at 0.5\% on the cross section; at strong cuts  $v_{max}\rightarrow 0$ we have $0.025\%$ and at moderate cuts near  $v_{max}\cong 0.6$ we have $.08\%$, as needed
for precision LHC studies. These estimates hold for both the IFI on and IFI off cases.

\def\Process{$u \bar{u} \to \mu^-\mu^+$}
\begin{table}[h!]
\centering
\includegraphics[width=90mm,height=60mm]{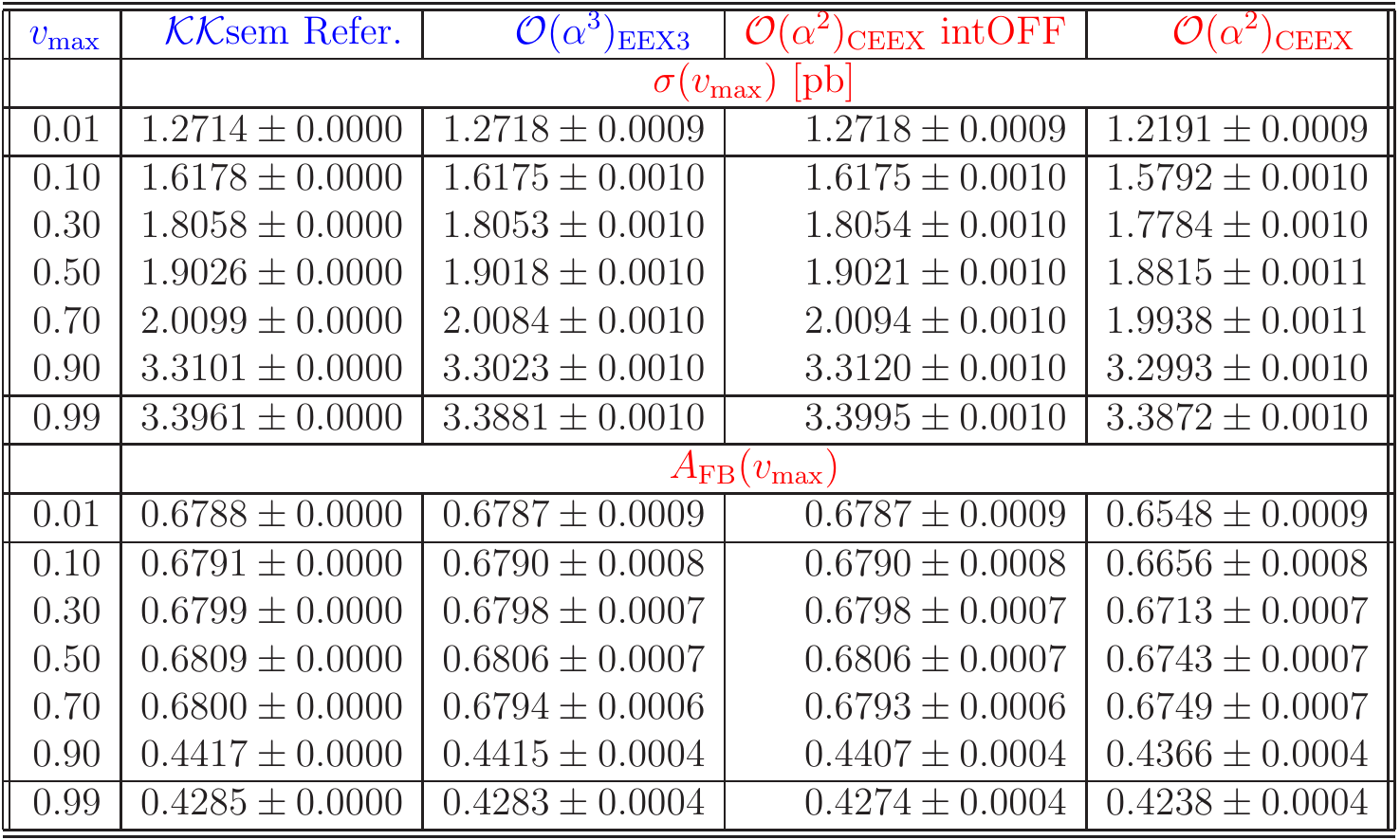}
\caption{
 Study of total cross section $\sigma(v_{\max})$ 
 and charge asymmetry $A_{\rm FB}(v_{\max})$,
 \Process, at $\sqrt{s}$~=\Energy.
 See Table \ref{tab:table1} for definition of
 the energy cut $v_{\max}$, scattering angle and M.E. type, 
}
\label{tab:table3}
\end{table}

\begin{figure}[h]
\includegraphics[width=71mm,height=40mm]{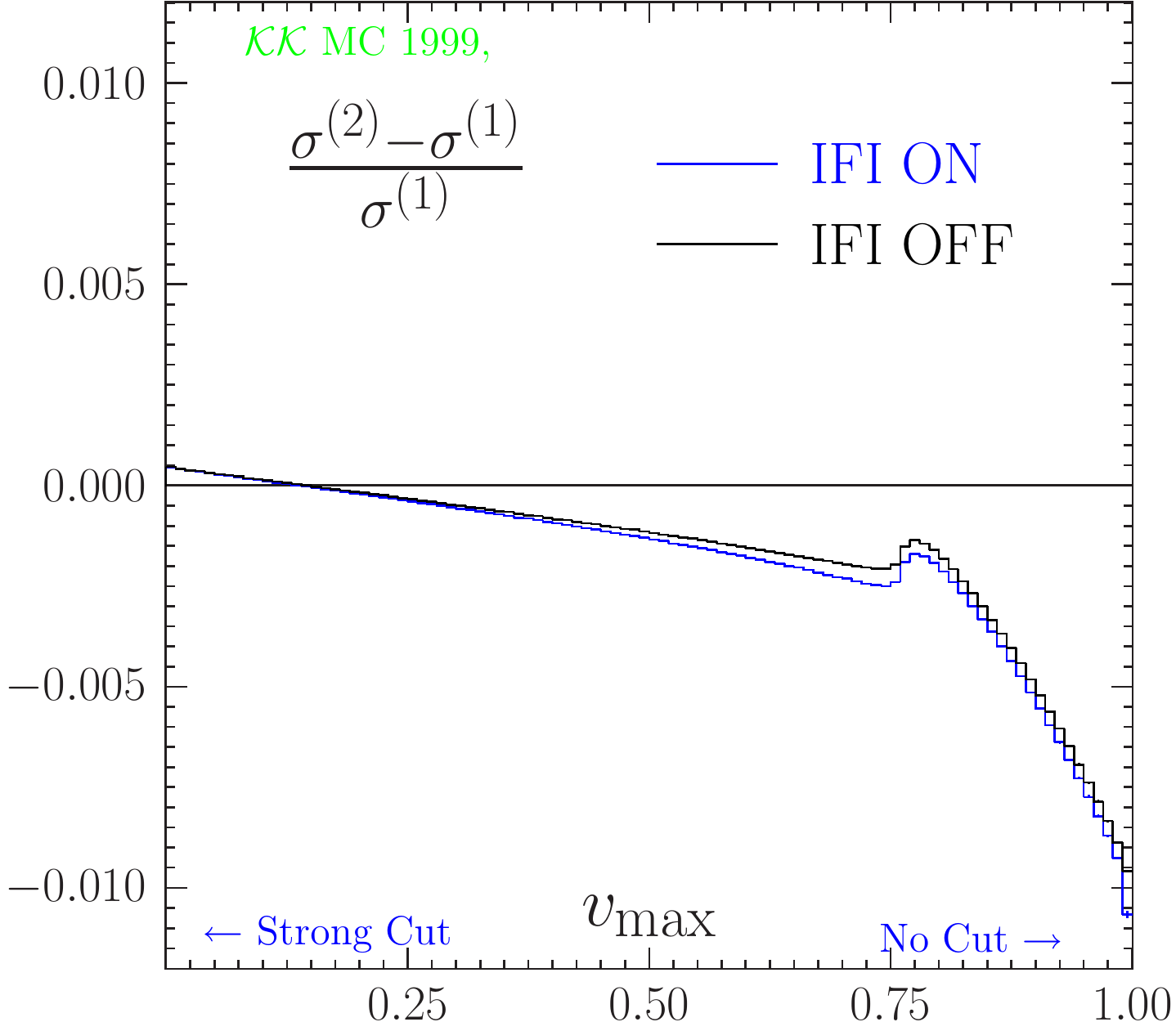}
\includegraphics[width=71mm,height=40mm]{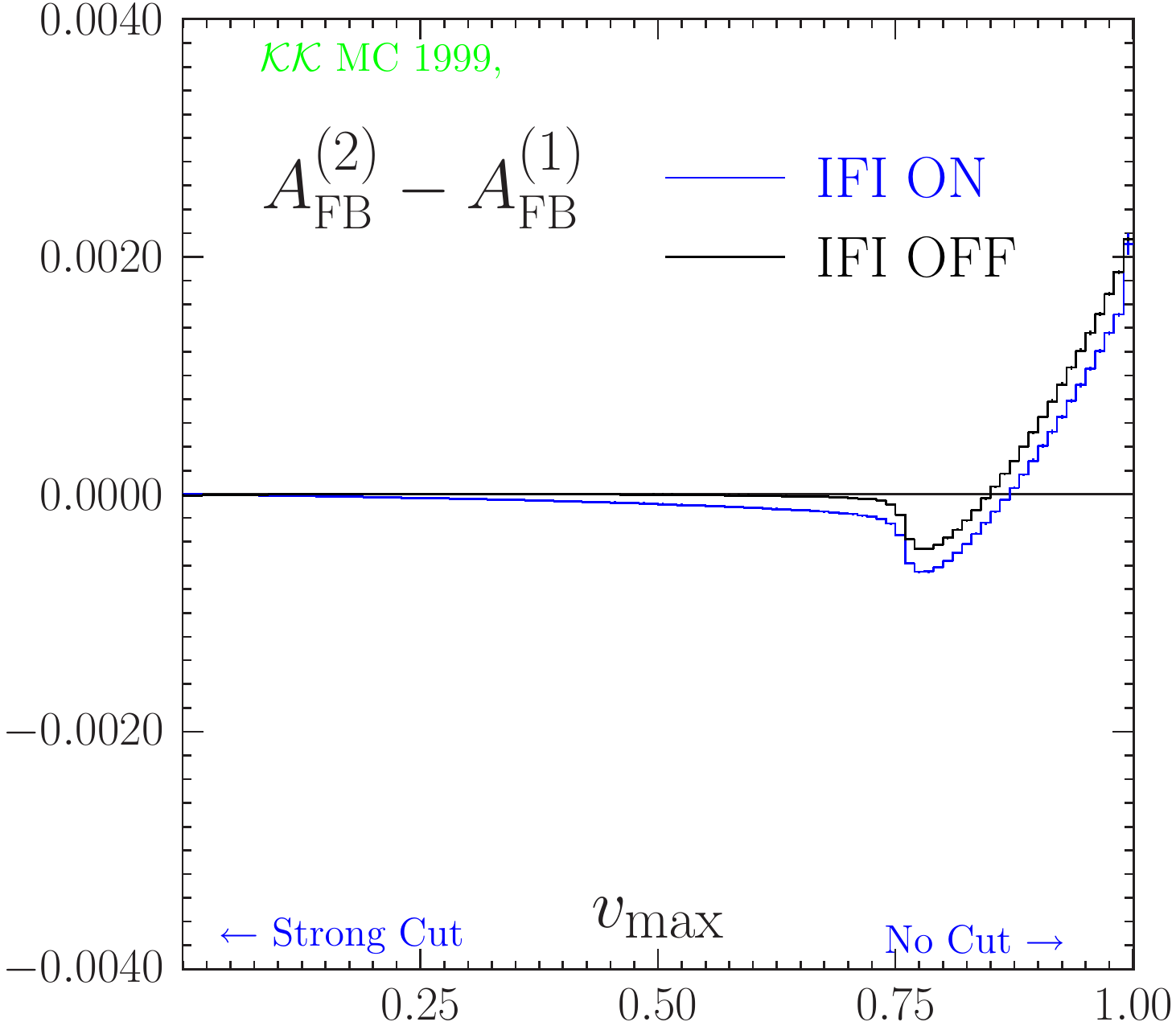}
\caption{
Physical precision of CEEX ISR matrix element
for \Process at $\sqrt{s}=$\Energy.
See table \ref{tab:table1} for definition of cut-offs.
}
\label{fig12}
\end{figure}

\begin{figure}[h]
\centering
\includegraphics[width=70mm]{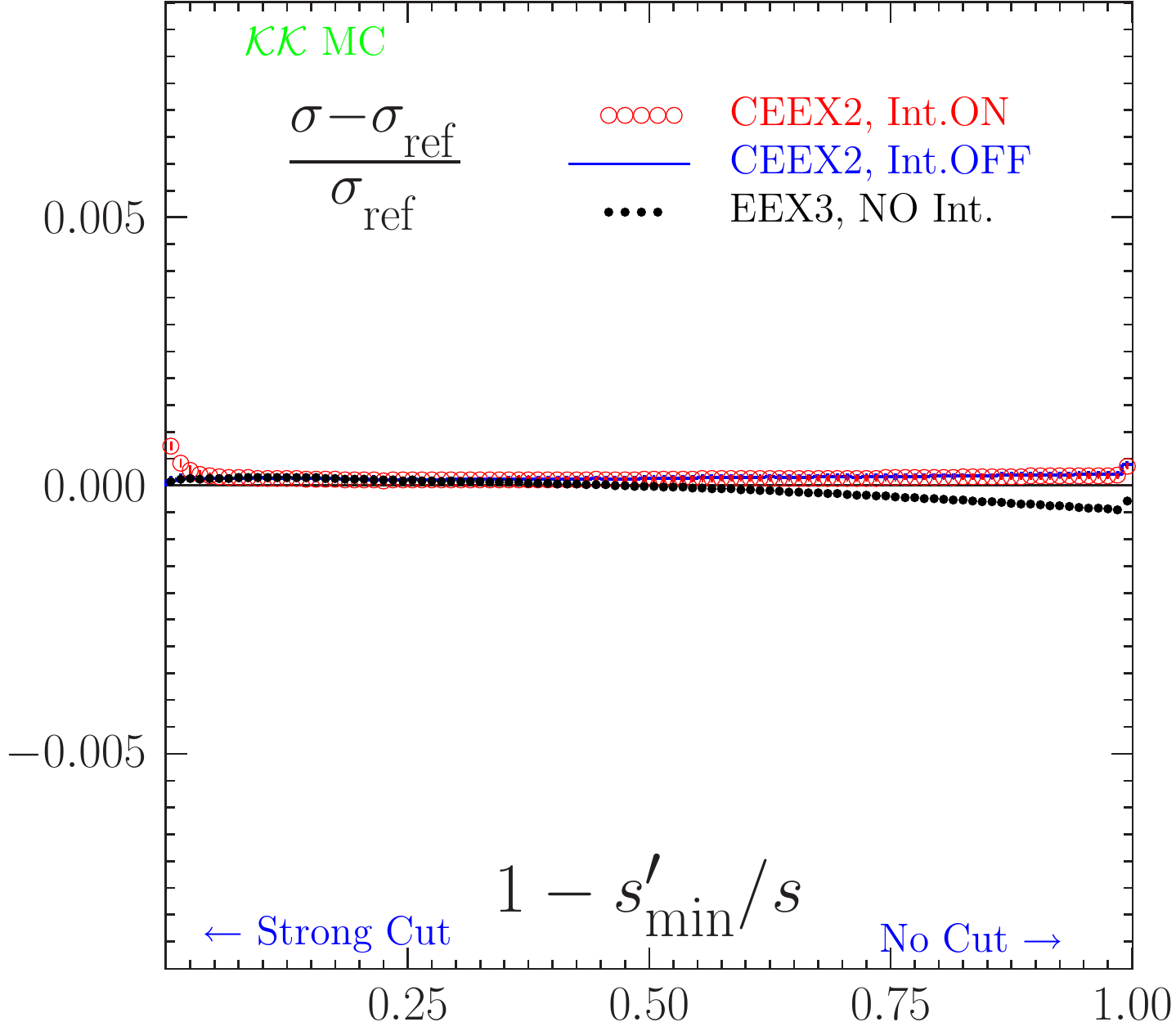}
\caption{
Total cross section $\sigma$, energy cut-off study for
the process $d\bar{d}\to\mu^-\mu^+$ at the $Z$.
Results the same as in the table \ref{tab:table4}.
}
\label{fig14}
\end{figure}

\begin{figure}[h]
\centering
\includegraphics[width=70mm]{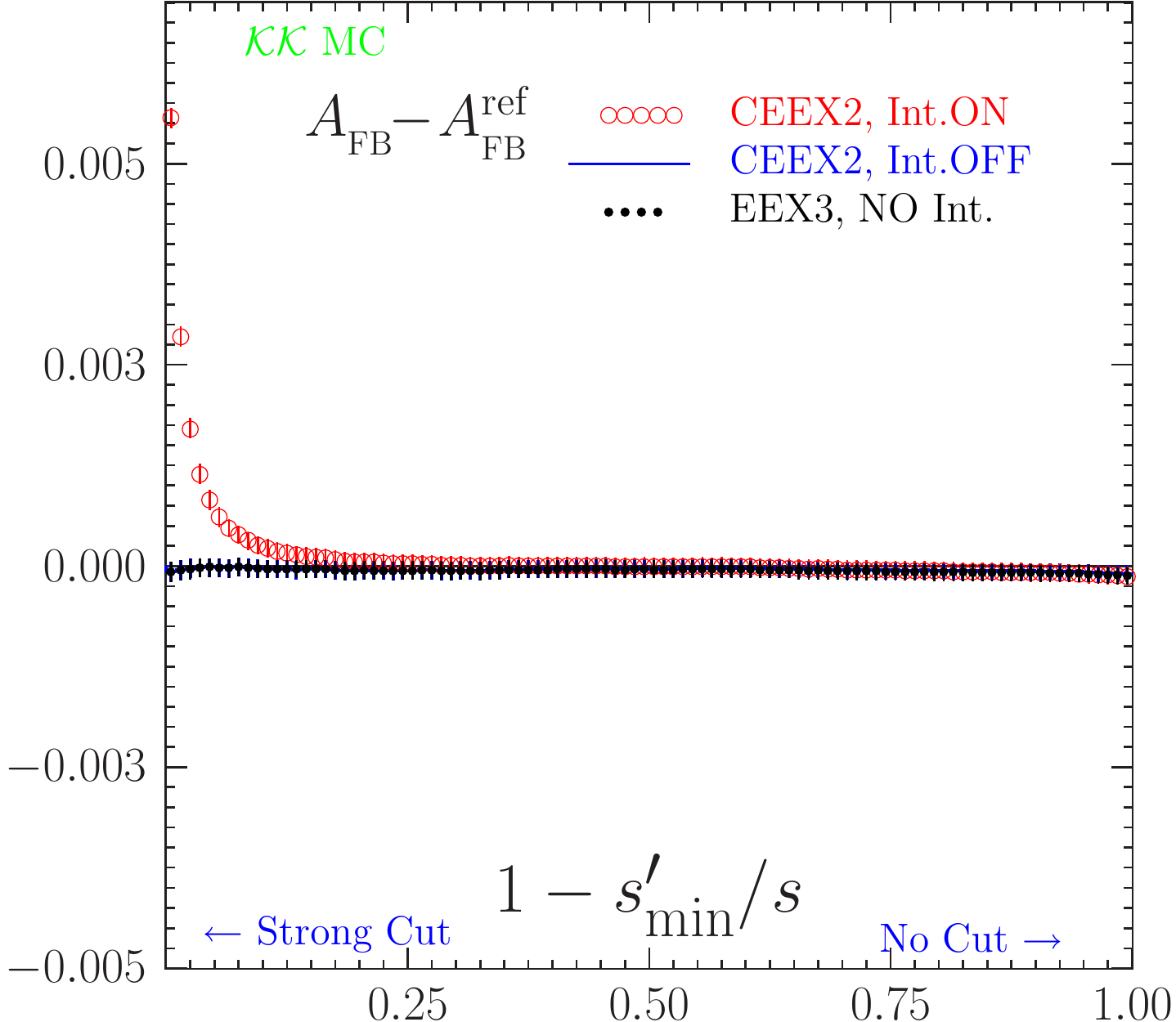}
\caption{
Charge asymmetry $A_{\rm FB}$, energy cut-off study for
the process $d\bar{d}\to\mu^-\mu^+$ at the $Z$.
Results the same as in the table \ref{tab:table4}.
}
\label{fig15}

\end{figure}

As most of the cross section at the LHC in the single $Z/\gamma*$ production and decay to lepton pairs is concentrated near the $Z-$resonance, we next turn to the similar studies as we have shown in Tabs.~\ref{tab:table1}-\ref{tab:table3} and Figs.~\ref{fig2}-\ref{fig12} for $\sqrt{s}=M_Z$ so see more directly what type of effects one has to consider in precision studies of these processes. We stress that with $25fb^{-1}$ of recorded data for each of ATLAS and CMS, the number of such decays exceeds 10 M in each experiment. Turning first to the $d\bar{d}$ incoming beam scenario we have the results in Tab.~\ref{tab:table4} and Figs.~\ref{fig14}-\ref{fig16}. We see that the small width(that is to say the lifetime) of the Z suppresses the IFI effects as expected: on the cross section even for the strong cut $v_{max}=0.01$ the effect is at the level of only $0.065\%$ and it is already essentially non-existent at $v_{max}=0.1$; on $A_{FB}$ a $5.5\%$ enhancement at $v_{max}=0.01$ is already reduced to $0.29\%$ at  $v_{max}=0.1$. But, the effect of the radiation on the cross section is quite pronounced, as the cross section changes by 26\% between the strong cut $v_{max}=0.01$ and the loose cut $v_{max}=0.99$. Thus, high precision on its theoretical prediction is essential for LHC precision studies. Indeed, these remarks are borne out in the plots in Figs.~\ref{fig14} and \ref{fig15}, where we respectively see the closeness of the CEEX cross section with the IFI on and IFI off and the similar closeness of the CEEX forward-backward asymmetries with the IFI on and off except for the region below $v_{max}=0.01$, where the IFI
effect reaches $5.5\%$. Turning to the physical precision study in Fig.~\ref{fig16}, we see that in the typical scenario where $v_{max}\cong 0.6$, the precision tag for both IFI on and the IFI off cross sections is $0.04\%$, sufficient for the precision LHC studies. 

\def\Energy{  91.187GeV}
\def\Process{$d \bar{d} \to \mu^-\mu^+$}
\begin{table}[h]
\centering
\includegraphics[width=90mm,height=60mm]{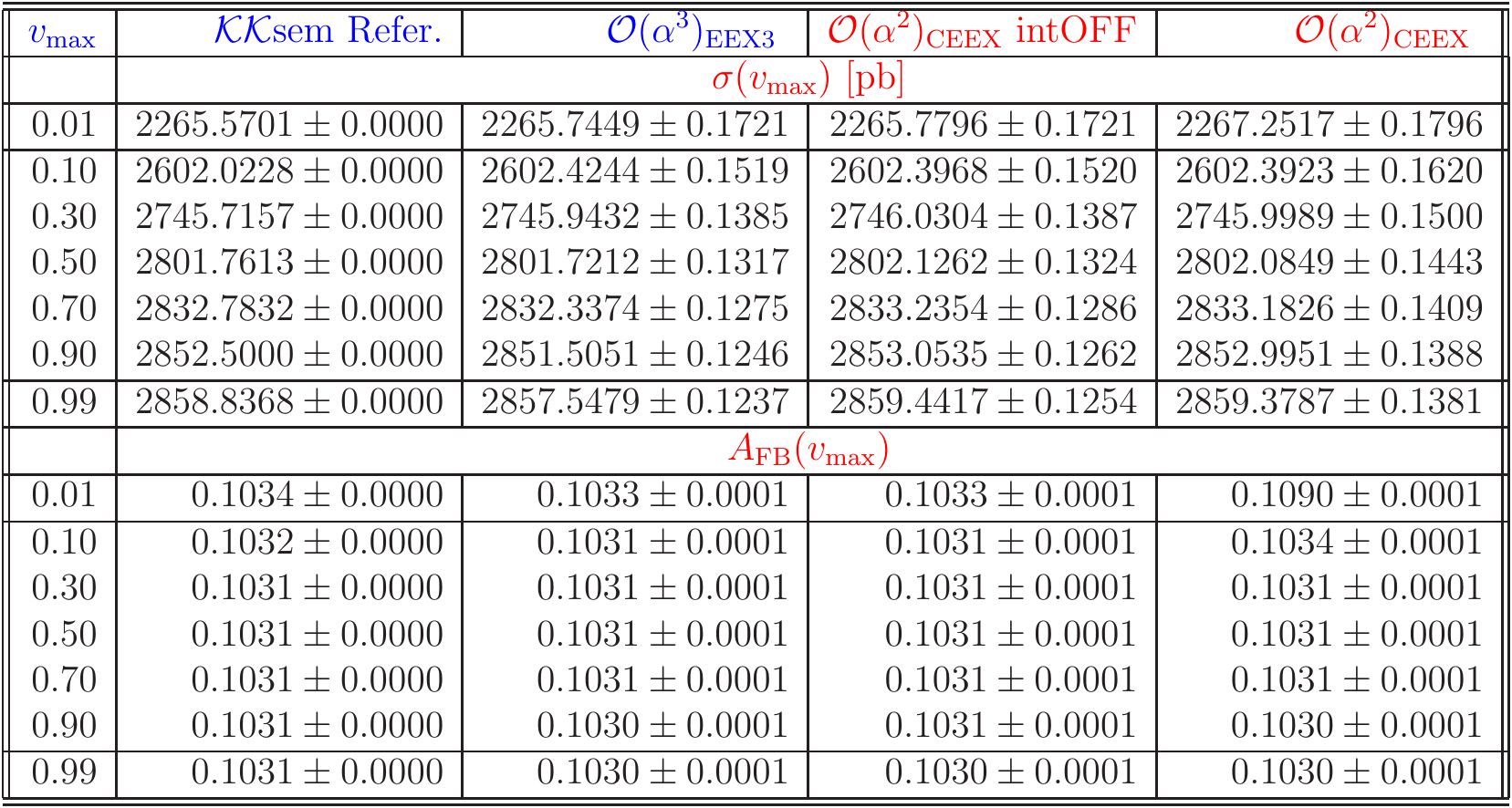}
\caption{
 Study of total cross section $\sigma(v_{\max})$ 
 and charge asymmetry $A_{\rm FB}(v_{\max})$,
 \Process, at $\sqrt{s}$~=\Energy.
 See Table \ref{tab:table1} for definition of
 the energy cut $v_{\max}$, scattering angle and M.E. type, 
}
\label{tab:table4}
\end{table}

\begin{figure}[h]
\centering
\includegraphics[width=71mm,height=40mm]{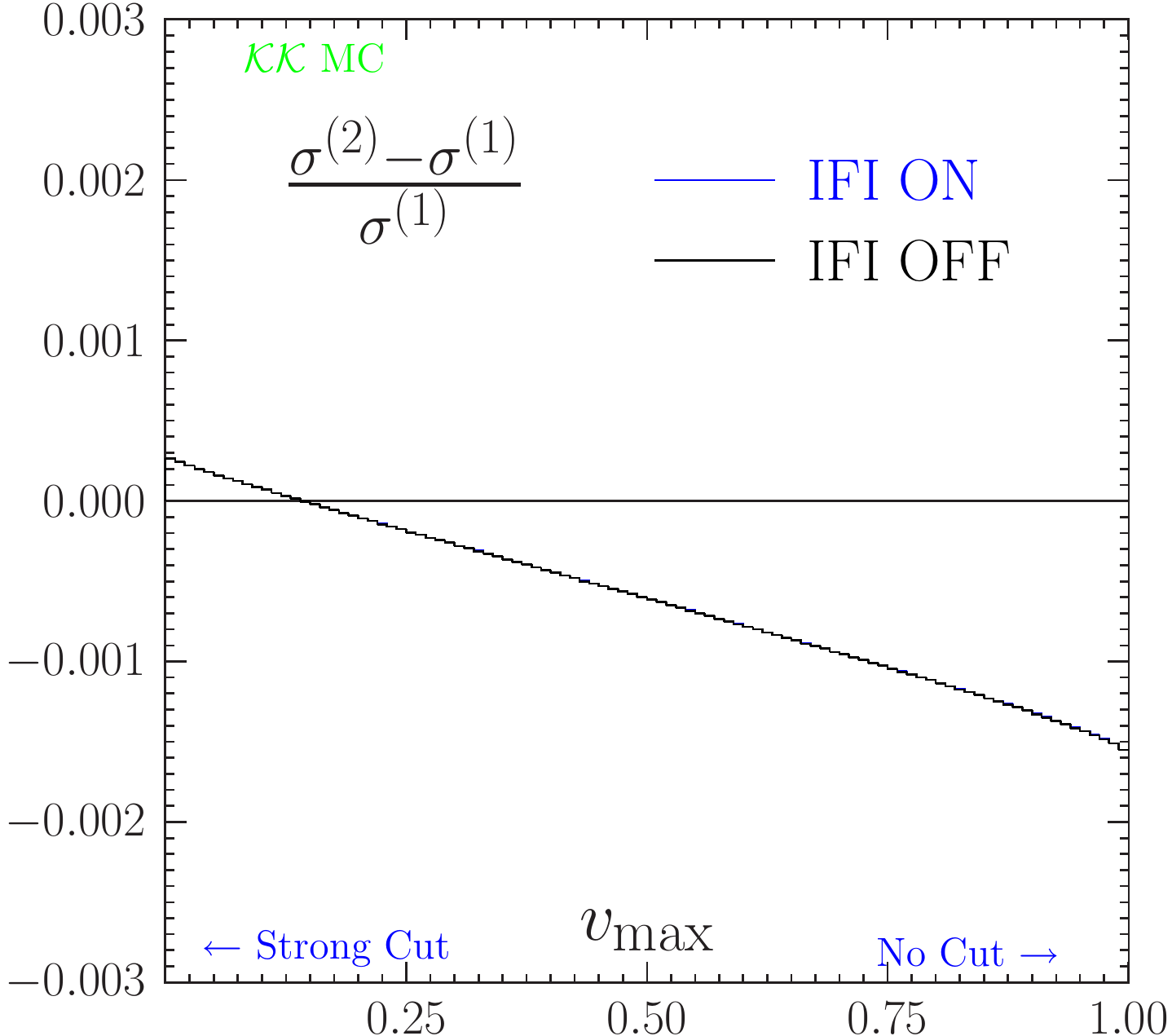}
\includegraphics[width=71mm,height=40mm]{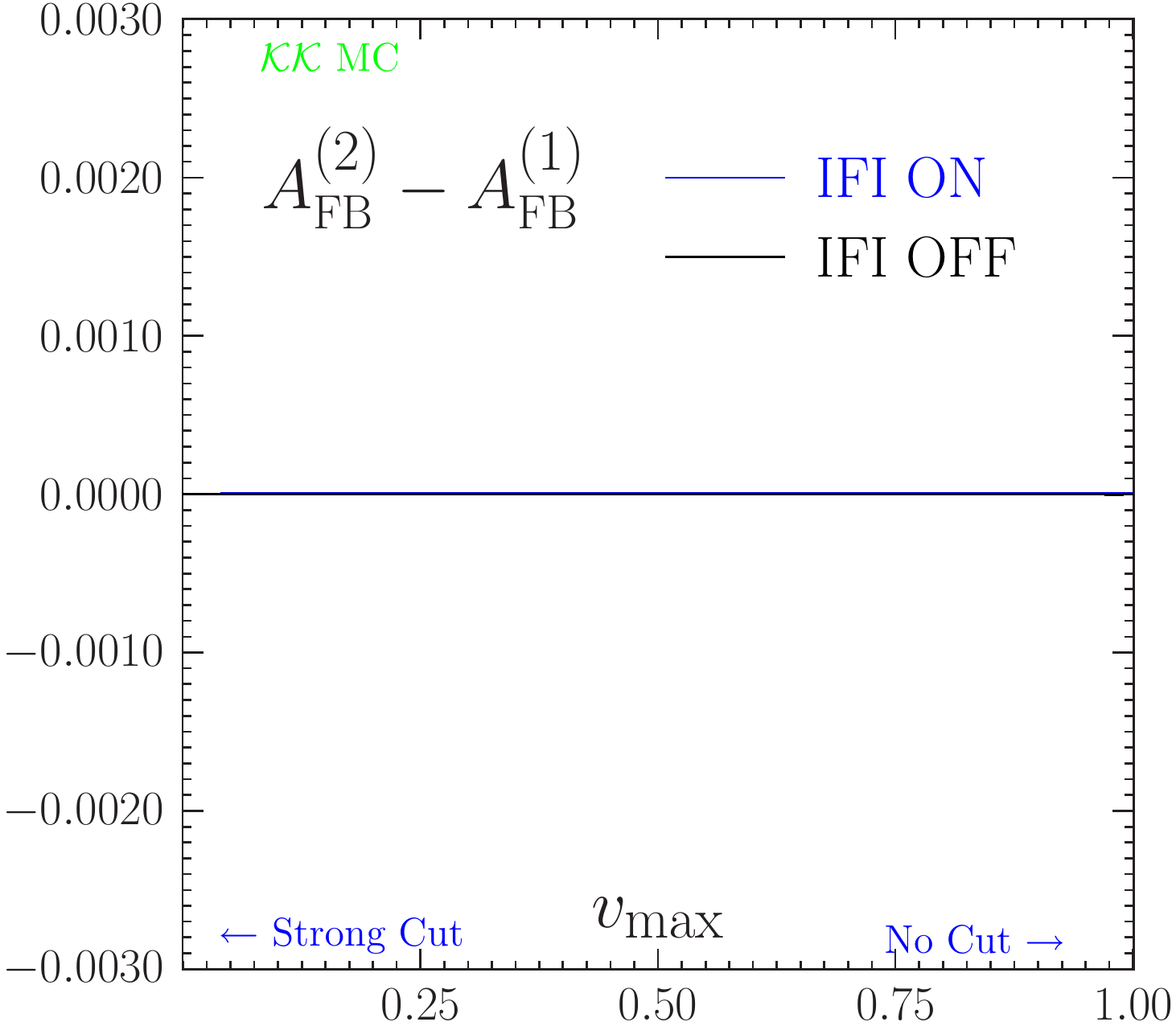}
\caption{
Physical precision of CEEX ISR matrix element
for \Process at $\sqrt{s}=$\Energy.
See table \ref{tab:table1} for definition of cut-offs.
}
\label{fig16}
\end{figure}

\begin{figure}[h]
\includegraphics[width=70mm]{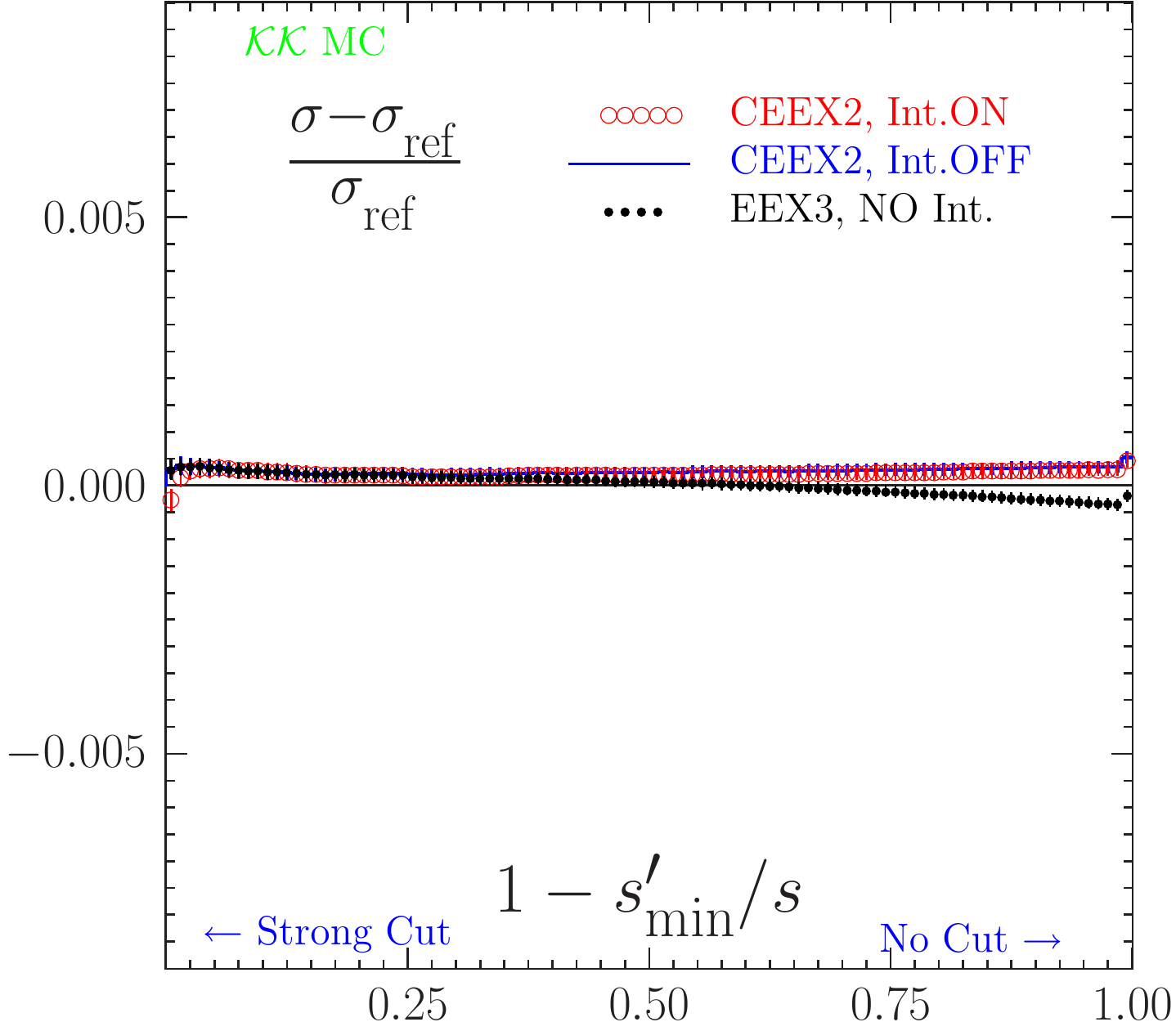}
\caption{
Total cross section $\sigma$, energy cut-off study for
the process $u\bar{u}\to\mu^+\mu^-$ at the $Z$ peak.
Results the same as in the table \ref{tab:table5}.
}
\label{fig18}
\end{figure}

\begin{figure}[h]
\centering
\includegraphics[width=70mm]{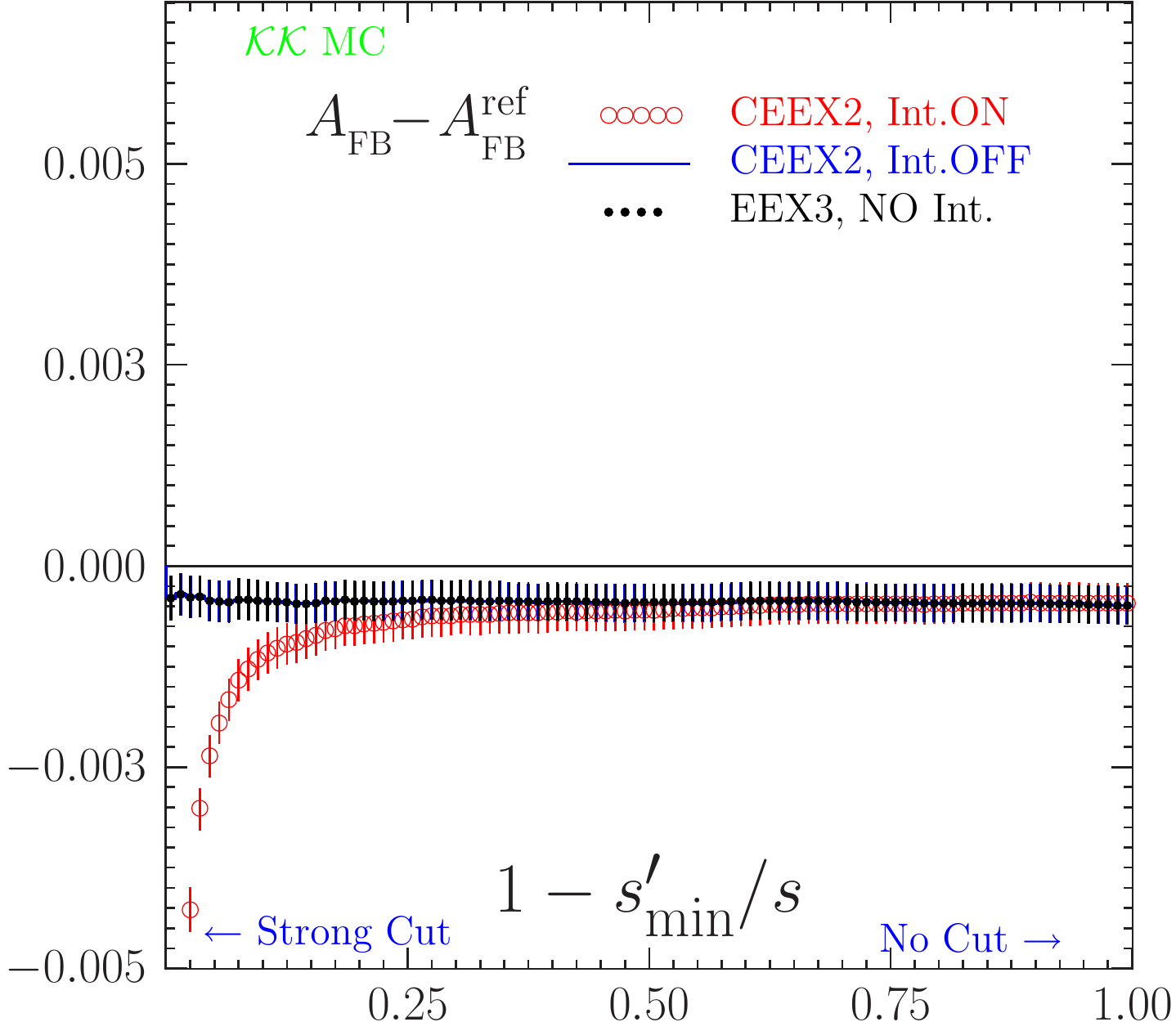}
\caption{
Charge asymmetry $A_{\rm FB}$, energy cut-off study for
the process $u\bar{u}\to\mu^-\mu^+$ at the $Z$.
Results the same as in the table \ref{tab:table5}.
}
\label{fig19}
\end{figure}

Continuing in this vein, we present next the incoming $u\bar{u}$ scenario at $\sqrt{s}=M_Z$ in Tab.~\ref{tab:table5} and Figs.~\ref{fig18}-\ref{fig20}. We see again that 
that the small width of the Z suppresses the IFI effects: the negative effects at $v_{max}=0.01$ of $-0.0587\%$ on the cross section and $-16.2\%$ on $A_{FB}$
become respectively non-existent and $-.989\%$ at  $v_{max}=0.1$; at the loose cut $v_{max}=0.99$ the IFI effect on the cross section(the forward-backward asymmetry) is below the $0.01\%(0.00285)$ precision of the data. The cross section varies by $30.6\%$ as $v_{max}$ varies from $0.01$ to $0.99$ so again its theoretical prediction for the radiative effects must have high precision for precision studies. These remarks are borne out by the plots in Figs.~\ref{fig18} and \ref{fig19}, where see that the IFI on and IFI CEEX cross sections are very close to the reference cross section even for the very strong and loose cuts and that 
the IFI on and off CEEX forward-backward asymmetries are the same as the EEX3
value by an energy cut value of $0.25$, for example. In Fig.~\ref{fig20}, we see the precision study shows that the cross section has the precision estimate of $0.04\%$ at the energy cut of $0.6$ just as we had for the incoming $d\bar{d}$
case. Again, this is sufficient for precision studies of LHC physics.
\def\Process{$u \bar{u} \to \mu^-\mu^+$}

\begin{table}[h]
\centering
\includegraphics[width=90mm,height=60mm]{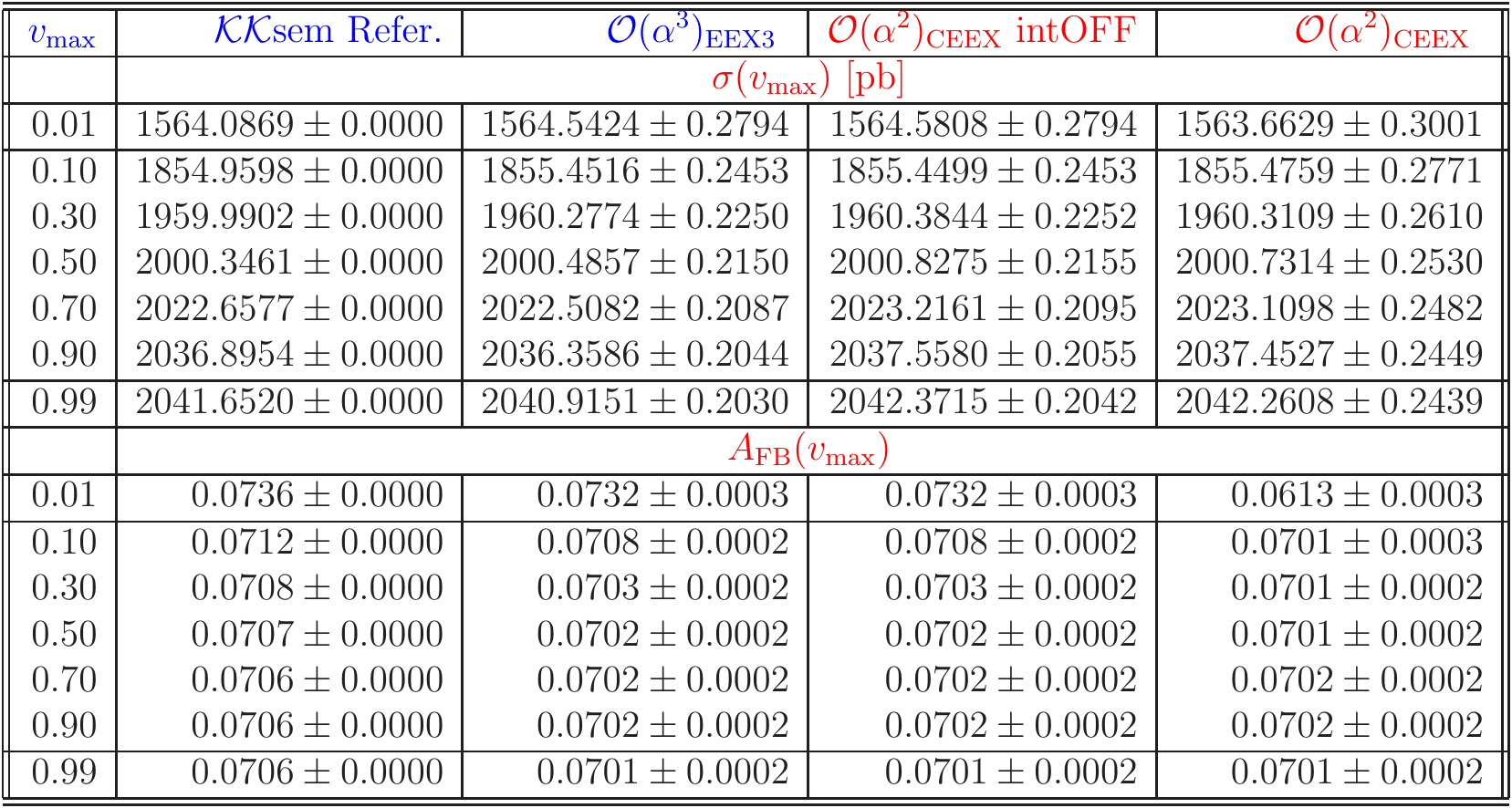}
\caption{
 Study of total cross section $\sigma(v_{\max})$ 
 and charge asymmetry $A_{\rm FB}(v_{\max})$,
 \Process, at $\sqrt{s}$~=\Energy.
 See Table \ref{tab:table1} for definition of
 the energy cut $v_{\max}$, scattering angle and M.E. type, 
}
\label{tab:table5}
\end{table}

\begin{figure}[h]
\centering
\includegraphics[width=71mm,height=40mm]{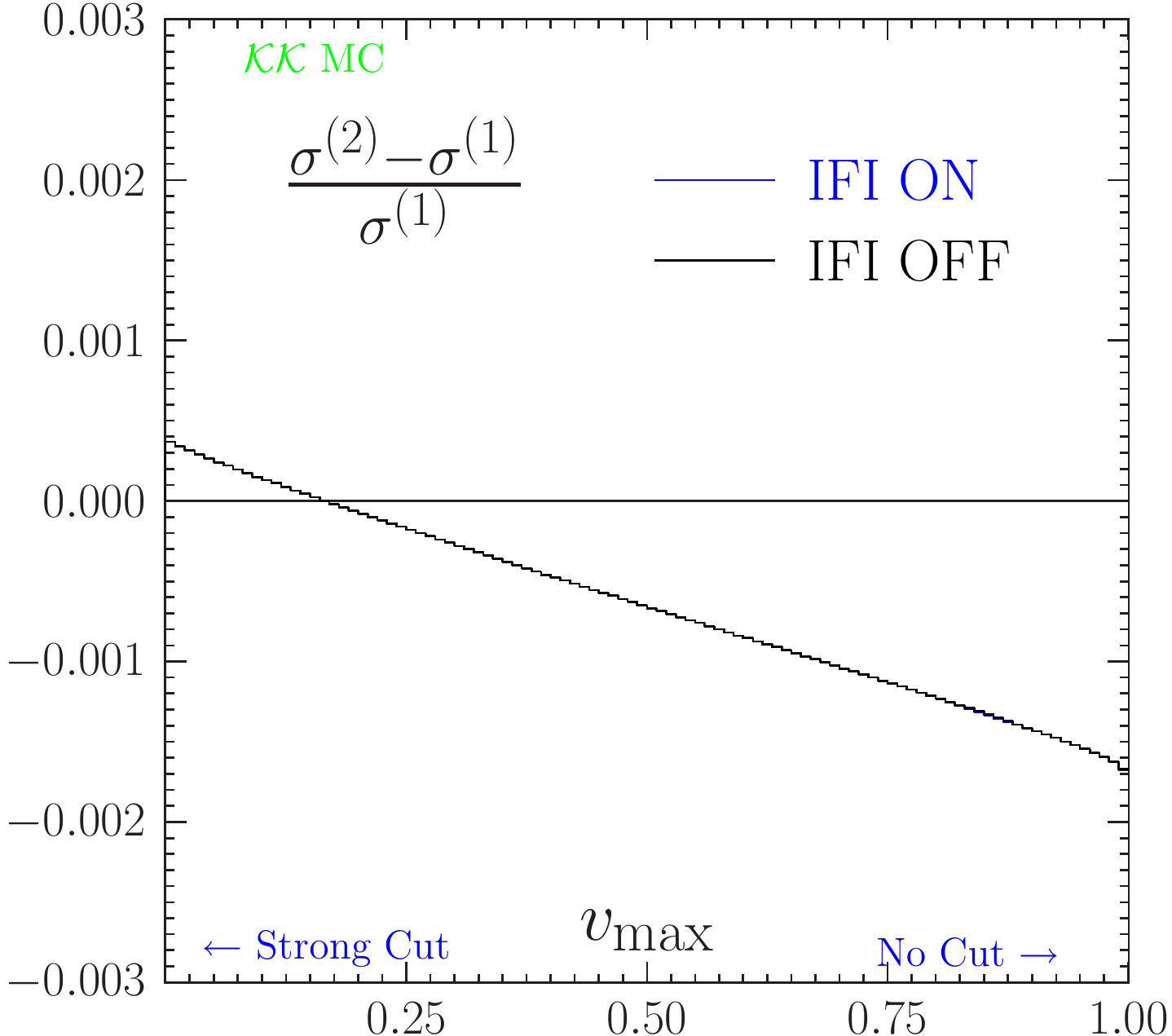}
\includegraphics[width=71mm,height=40mm]{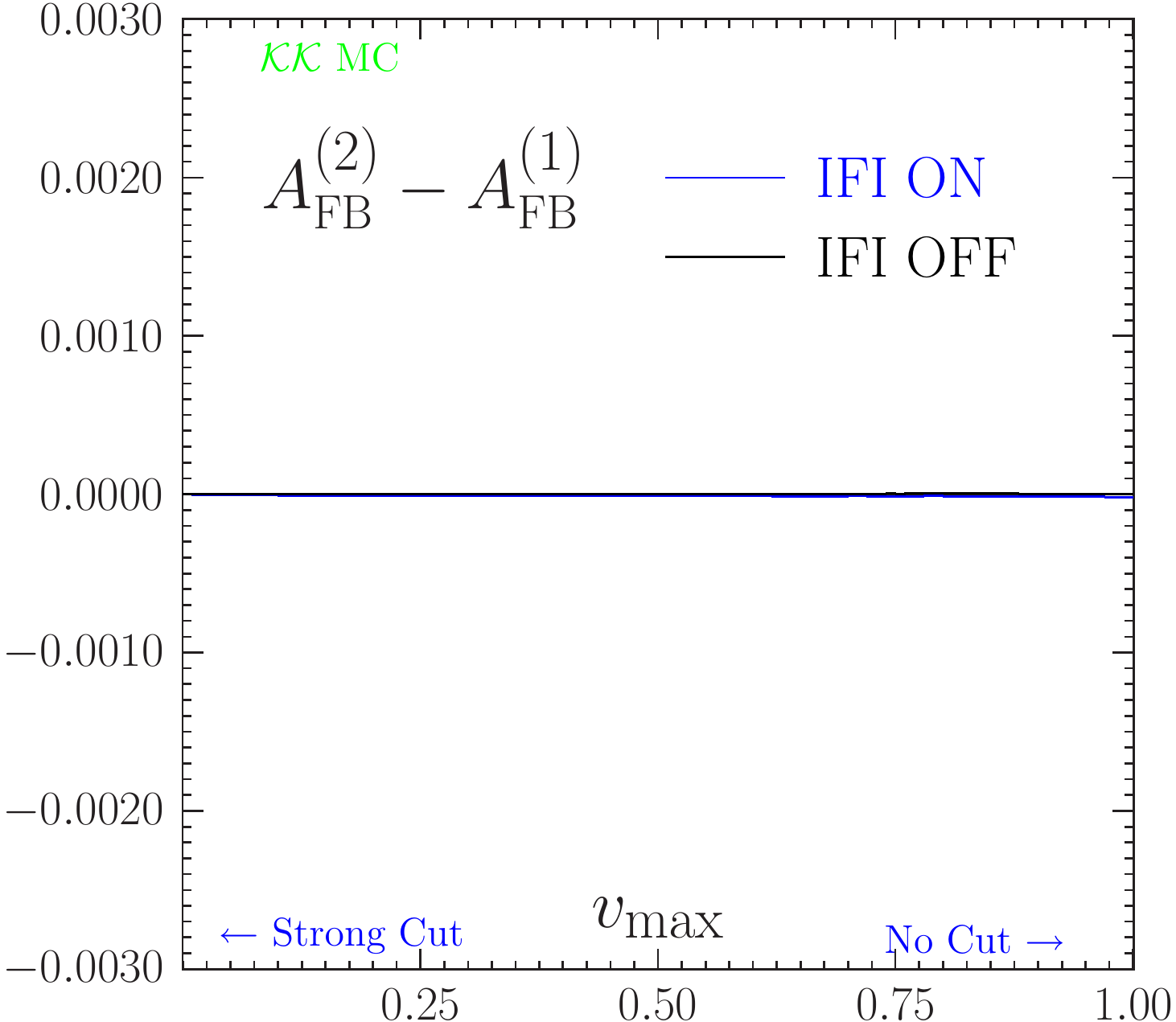}
\caption{
Physical precision of CEEX ISR matrix element
for \Process at $\sqrt{s}=$\Energy ($Z$ peak).
See table \ref{tab:table1} for definition of cut-offs.
}
\label{fig20}
\end{figure}

While we have discussed the individual incoming $q\bar{q}$ scenarios, 
\KK MC 4.22
has a beamstrahlung option in which one may replace the beamstrahlung functions with the proton PDF's. We have done this as a proof of principle exercise and we show in Appendix 1 the results of a simple test run at $7$TeV. What we see in this test run output is that indeed significant probability exists for the incoming quarks to radiate non-zero $p_T$ in the higher order corrections: these effects cannot be properly described by zero $p_T$ methods such as structure function techniques~\cite{hor}. We will return to such studies elsewhere~\cite{elsewh}.
\par
Finally, given the interest in muon collider precision physics~\cite{muclldr}, we consider next the process $\mu^+\mu^-\rightarrow e^+e^-$ again 
at $\sqrt{s}=189$GeV, so that again 
we have the reference to the usual
incoming $e^+e^-$ annihilation case regarding the size and nature of the 
EW effects expected. In this case we have all the same EW charges but the ISR probability to radiate factor $\gamma_e=\frac{2\alpha}{\pi}\left(\ln(s/m_e^2)-1\right)\cong 0.114$ becomes $\gamma_\mu=\frac{2\alpha}{\pi}\left(\ln(s/m_\mu^2)-1\right)\cong 0.0649$. This means that we expect the EW effects where the photonic corrections dominate to show reduction in size for ISR dominated regimes, the same size for the IFI dominated regimes. This is borne-out by the results in Tab.~\ref{tab:table6} and Figs.~\ref{fig22}-\ref{fig24}. In the regime of the strong cut, 
with $v_{max}=0.01$, the results
are very similar in all aspects to the usual incoming $e^-e^+$ case: the cross section is enhanced by $6.0\%$ to be compared with $5.9\%$ and $A_{FB}$ is enhanced by $8.3\%$ to be compared to $8.1\%$. In the regime of the loose cut, with $v_{max}=0.99$, the cross section is enhanced by $0.49\%$ to be compared 
with $0.38\%$
and $A_{FB}$ is enhanced by $1.7\%$ to be compared to $2.4\%$. In Figs.~\ref{fig22} and \ref{fig23} we see that we have same general behavior as we have in Figs.~\ref{fig2} and \ref{fig3}, the characteristic Z peak radiative return structure
in Fig.~\ref{fig22} and its inflection behavior in Fig.~\ref{fig23}. In Fig.~\ref{fig24}, we see that the precision studies comparing the second order and first order CEEX results show the pronounced effect of the Z radiative return. At an energy cut of $0.6$, we see again that a precision tag of $0.2\%$  obtains, so that precision results for EW effects would be available. The detailed application of such results to muon collider physics will be taken up elsewhere~\cite{elsewh2}.

\def\Energy{ 189GeV}
\def\Process{$ \mu^- \mu^+ \to e^- e^+$}
\begin{table}[h]
\centering
\includegraphics[width=90mm,height=60mm]{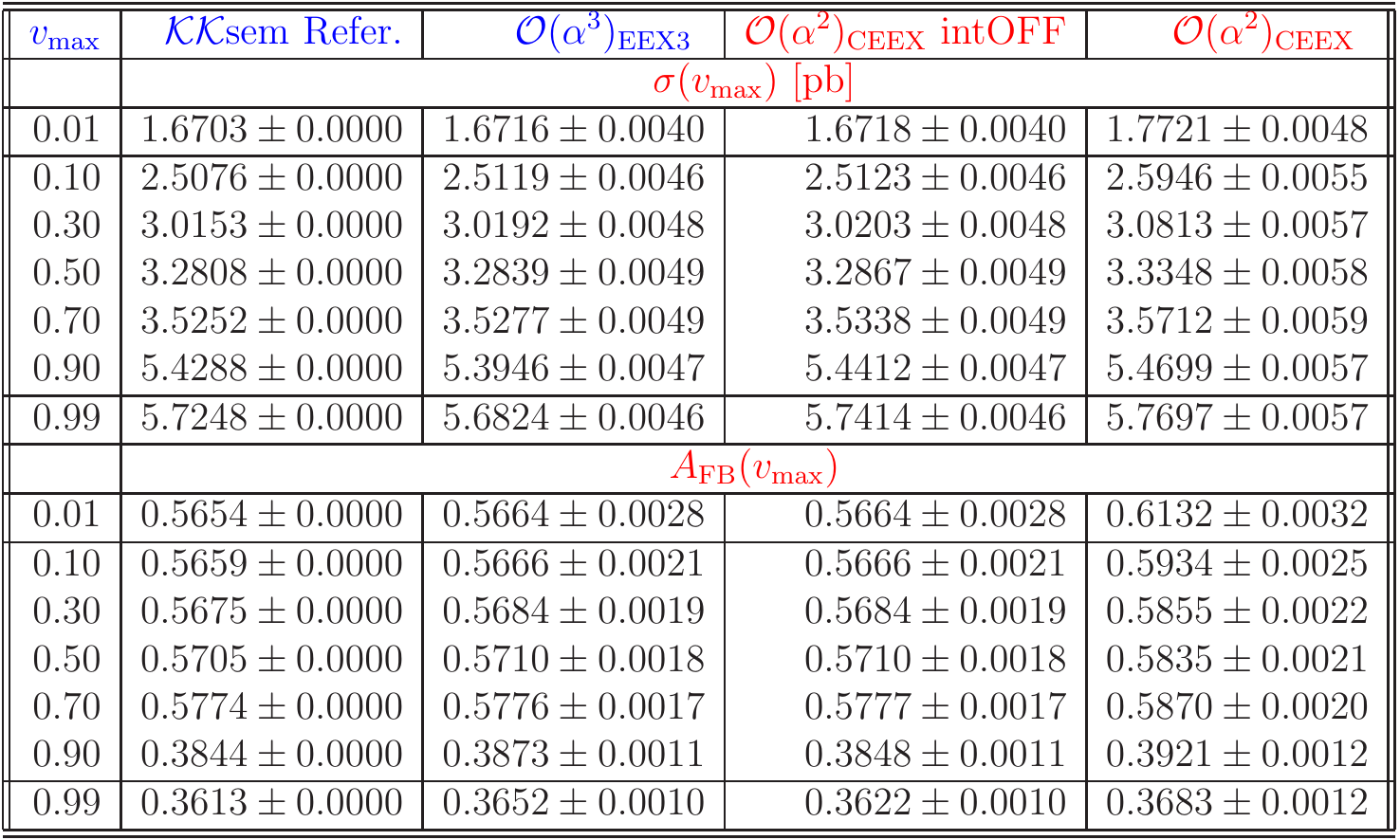}
\caption{
 Study of total cross section $\sigma(v_{\max})$ 
 and charge asymmetry $A_{\rm FB}(v_{\max})$,
 \Process, at $\sqrt{s}$~=\Energy.
 See Table \ref{tab:table1} for definition of
 the energy cut $v_{\max}$, scattering angle and M.E. type, 
}
\label{tab:table6}
\end{table}

\begin{figure}[h]
\centering
\includegraphics[width=70mm]{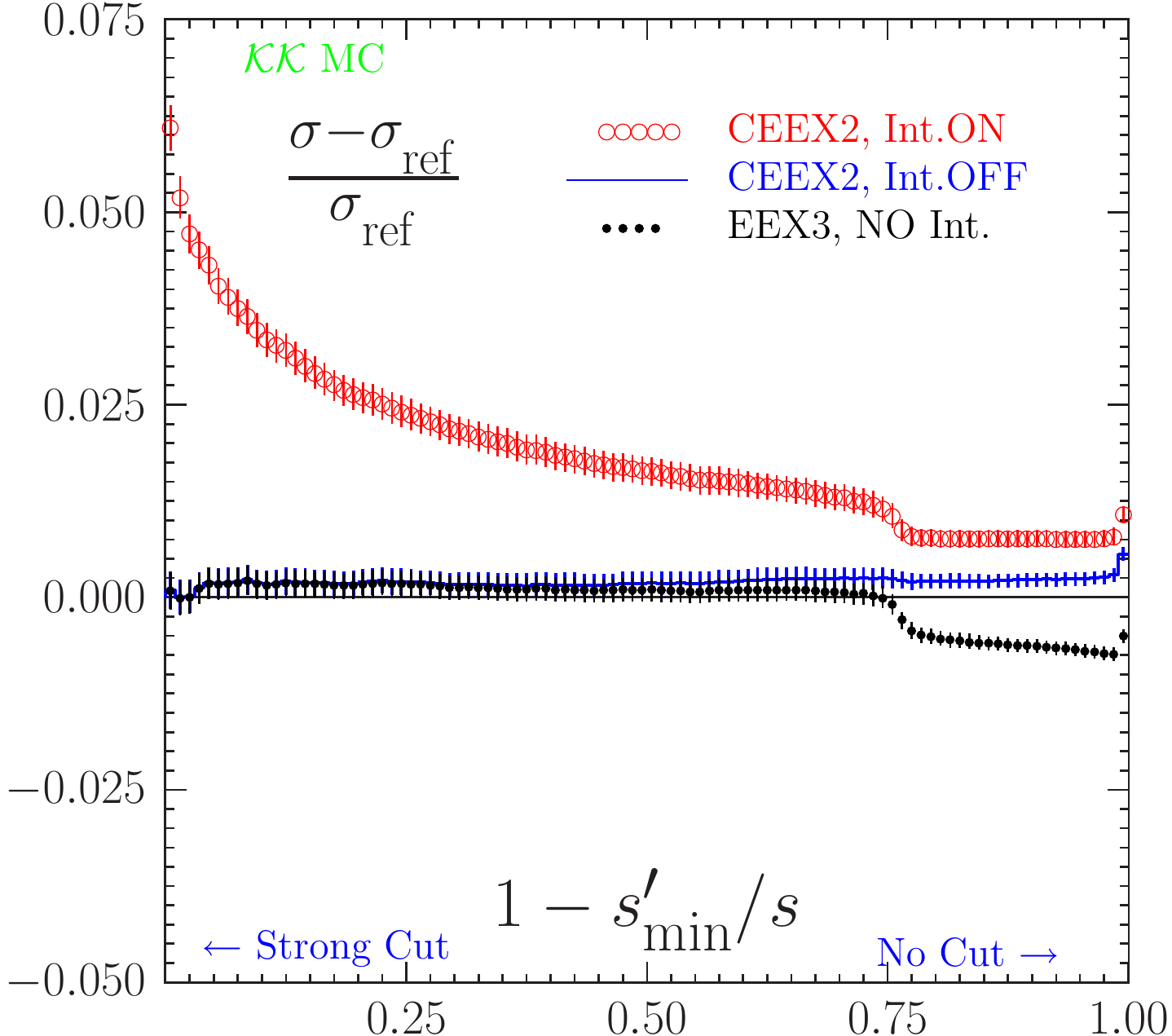}
\caption{
Energy cut-off study of total cross section
for \Process at energy \Energy.
The same results as in Table \ref{tab:table6}.
}
\label{fig22}
\end{figure}

\begin{figure}[h]
\centering
\includegraphics[width=70mm]{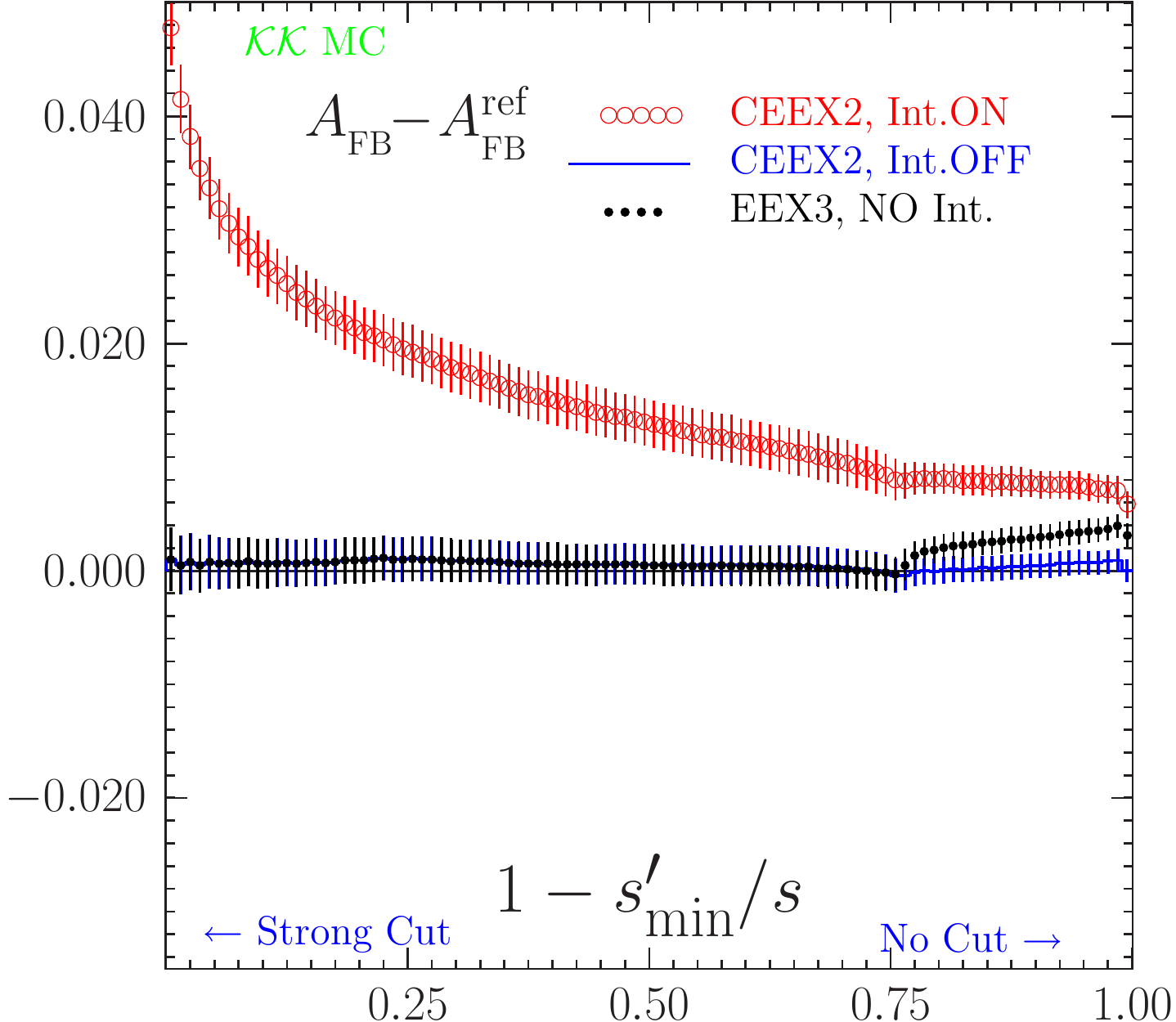}
\caption{
Energy cut-off study of total charge asymmetry
for \Process at energy \Energy.
The same results as in Table \ref{tab:table6}.
}
\label{fig23}
\end{figure}

\begin{figure}[h]
\centering
\includegraphics[width=71mm,height=40mm]{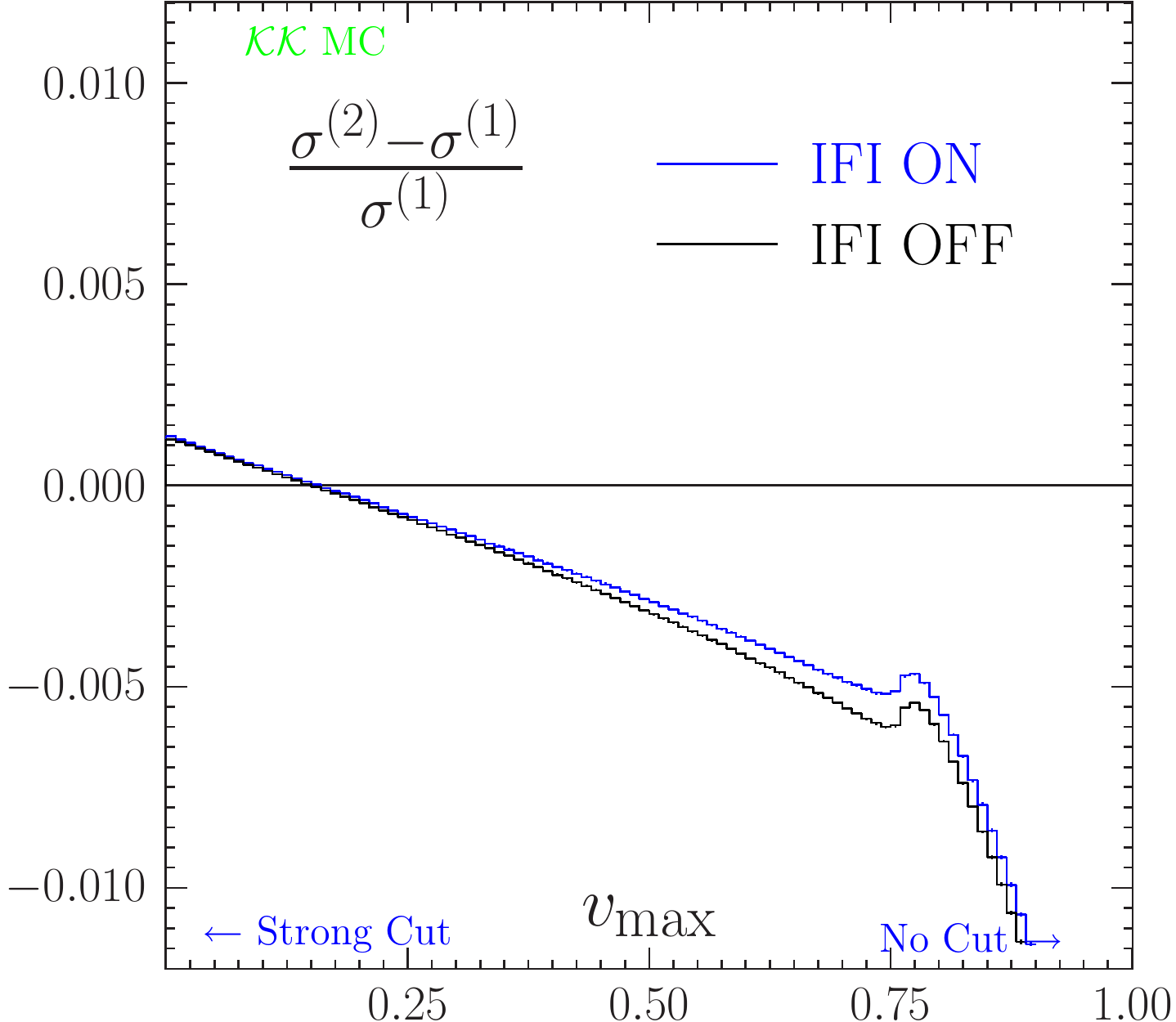}\\
\includegraphics[width=71mm,height=40mm]{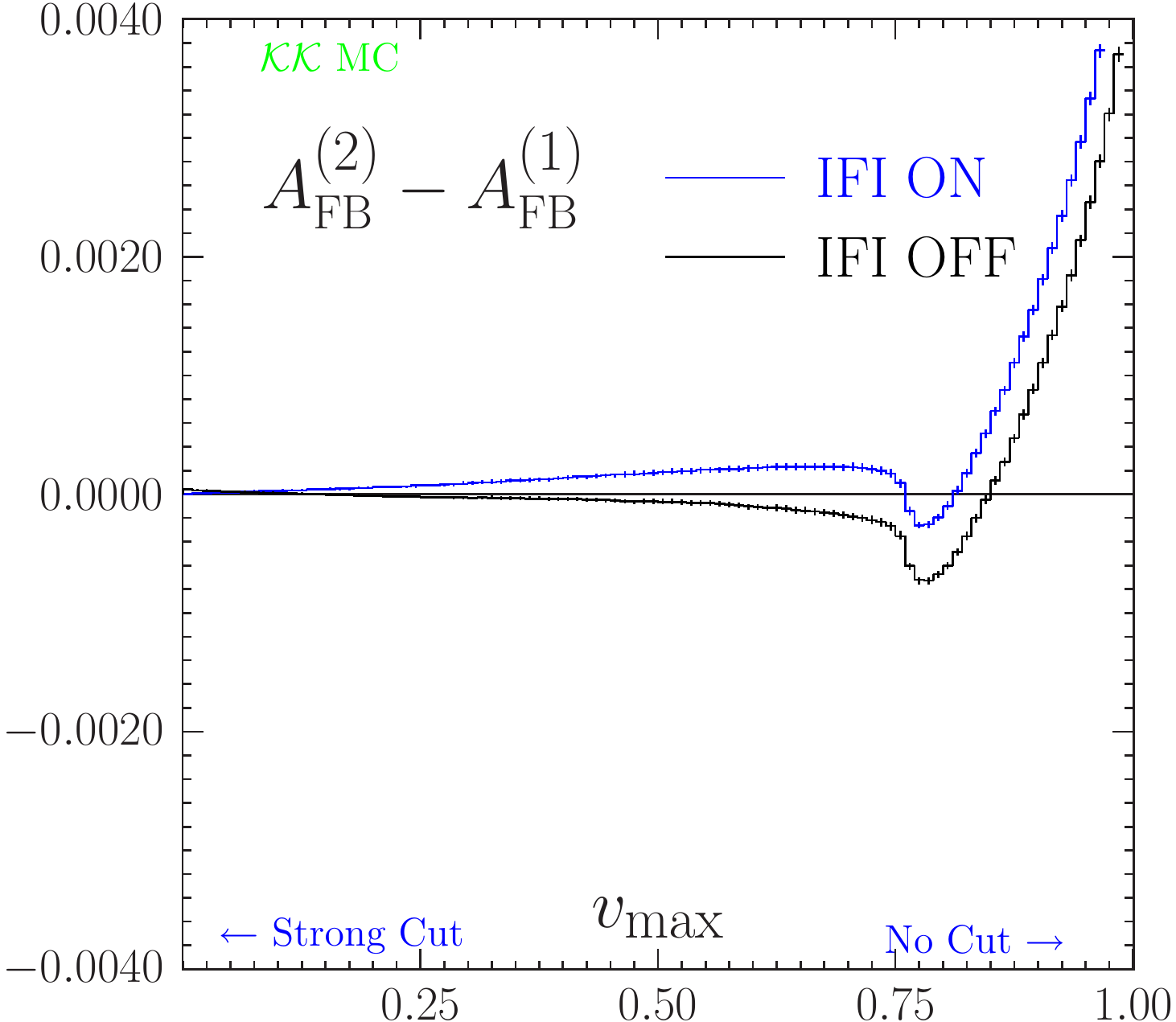}
\caption{
Physical precision of CEEX ISR matrix element
for \Process at $\sqrt{s}=$\Energy.
See table \ref{tab:table1} for definition of cut-offs.
}
\label{fig24}
\end{figure}

\section{Conclusions}
 
YFS inspired EEX and CEEX MC schemes are successful examples
of Monte Carlos based directly on the factorization theorem
(albeit for the IR soft case for Abelian QED only). These schemes
work well in practice: KORALZ, BHLUMI, YWSWW3, BHWIDE and \KK MC are
examples. The extension of such schemes (as far as possible) to 
all collinear singularities
would be very desirable and practically important! Work on this
is in progress-- see Refs.~\cite{sjms,herwiri1,herwiri2} for recent results
and outlooks.\par

Here, we have illustrated that the \KK MC program is extended to the 
new incoming $f\bar{f},\; f=\mu,\tau,\nu_\ell,q, \; q=u,d,s,c,b,\; \ell= e,\mu,\tau,$ beams cases. The quark-anti-quark and $\mu^-\mu^+$ incoming beam cases 
are respectively important for the LHC
precision EW predictions at the per mille level and to the
precision EW studies for the possible muon collider physics program.
We have seen that in all cases, the per mille level accuracy requirements
necessitate the implementation of the \KK MC class of EW higher order effects.
Realizations and applications of this class of higher order EW effects is in progress and will appear elsewhere~\cite{elsewh}.
The new version of the \KK MC, version 4.22, is available at
{https://jadach.web.cern.ch/jadach/KKindex.html}

\vspace{14mm}
\noindent
{\bf Acknowledgments}\\
The authors thank Prof. I. Antoniadis 
for the support and kind hospitality of the CERN Theory Division 
while this work was in progress. 
They also thank Dr. S.A. Yost for useful discussions.
One of the authors (S.J.) also thanks 
the Dean Lee Nordt of the Baylor College of Arts \& Sciences 
for Baylor's support while this work was in progress.
This work is partly supported by
the Polish National Science Centre grant DEC-2011/03/B/ST2/02632.

\appendix

\section{Sample Monte Carlo events}
\label{app:A}

Below sample output from run of \KK\ MC version 4.22 is presented
for $pp\to u\bar{u} \to l^- l^+ +n\gamma$ where
simple parton distribution functions (PDF's)
of $u$ and $\bar{u}$ quarks
in the proton are replacing beamstrahlung distributions
(see function {\tt BornV\_RhoFoamC} in the source code).
Three events are shown in the popular LUND MC format.
Two photons in the event record with the 
exactly zero transverse momentum,
formerly beamstrahlung photons, are now representing 
proton remnants (temporary fix).
What is important to see is the perfect energy momentum conservation
and proper flavor structure.
Overall normalization of the cross section is in principle 
also under strict control, however, more tests are needed.

\begin{widetext}
{\small
\baselineskip=3pt
\begin{verbatim}
 ***************************************************************************
 *                         KK Monte Carlo                                  *
 *            Version       4.22          May 2013                         *
 *    7000.00000000                 CMS energy average       CMSene     a1 *
 *       0.00000000                 Beam energy spread       DelEne     a2 *
 *              100                 Max. photon mult.        npmax      a3 *
 *                0                 wt-ed or wt=1 evts.      KeyWgt     a4 *
 *                1                 ISR switch               KeyISR     a4 *
 *                1                 FSR switch               KeyFSR     a5 *
 *                2                 ISR/FSR interferenc      KeyINT     a6 *
 *                1                 New exponentiation       KeyGPS     a7 *
 *                0                 Hadroniz.  switch        KeyHad     a7 *
 *       0.20000000                 Hadroniz. min. mass      HadMin     a9 *
 *       1.00000000                 Maximum weight           WTmax     a10 *
 *              100                 Max. photon mult.        npmax     a11 *
 *                2                 Beam ident               KFini     a12 *
 *       0.03500000                 Manimum phot. ener.      Ene       a13 *
 *   0.10000000E-59                 Phot.mass, IR regul      MasPho    a14 *
 *    1.2500000                     Phot. mult. enhanc.      Xenph     a15 *
 *       0.00000000                    PolBeam1(1)           Pol1x     a17 *
 *       0.00000000                    PolBeam1(2)           Pol1y     a18 *
 *       0.00000000                    PolBeam1(3)           Pol1z     a19 *
 *       0.00000000                    PolBeam2(1)           Pol2x     a20 *
 *       0.00000000                    PolBeam2(2)           Pol2y     a21 *
 *       0.00000000                    PolBeam2(3)           Pol2z     a22 *
 ***************************************************************************
  
                            Event listing (summary)
    I particle/jet KS     KF  orig    p_x      p_y      p_z       E        m
    1 !u!          21       2    0    0.000    0.000   22.668   22.668    0.005
    2 !ubar!       21      -2    0    0.000    0.000 -245.458  245.458    0.005
    3 (Z0)         11      23    1   23.016   18.370  -80.068  115.249   77.487
    4 gamma         1      22    1  -30.989   -6.132 -128.905  132.719    0.000
    5 gamma         1      22    1    0.000    0.000    0.031    0.031    0.000
    6 gamma         1      22    1    7.973  -12.238  -13.848   20.127    0.000
    7 gamma         1      22    1    0.000    0.000 3477.332 3477.332    0.000
    8 gamma         1      22    1    0.000    0.000-3254.542 3254.542    0.000
    9 tau-          1      15    3  -24.701   21.657  -20.217   38.613    1.777
   10 tau+          1     -15    3   47.716   -3.287  -59.851   76.635    1.777
                   sum:  0.00         0.000    0.000    0.000 7000.000 7000.000

                            Event listing (summary)
    I particle/jet KS     KF  orig    p_x      p_y      p_z       E        m
    1 !u!          21       2    0    0.000    0.000  271.908  271.908    0.005
    2 !ubar!       21      -2    0    0.000    0.000   -6.542    6.542    0.005
    3 (Z0)         11      23    1    0.047    1.133  244.401  257.454   80.928
    4 gamma         1      22    1   -0.047   -1.133   20.965   20.996    0.000
    5 gamma         1      22    1    0.000    0.000 3228.092 3228.092    0.000
    6 gamma         1      22    1    0.000    0.000-3493.458 3493.458    0.000
    7 mu-           1      13    3    0.601   14.537    2.005   14.687    0.106
    8 mu+           1     -13    3   -0.554  -13.404  242.396  242.767    0.106
                   sum:  0.00         0.000    0.000    0.000 7000.000 7000.000

                            Event listing (summary)

    I particle/jet KS     KF  orig    p_x      p_y      p_z       E        m
    1 !u!          21       2    0    0.000    0.000 1816.851 1816.851    0.005
    2 !ubar!       21      -2    0    0.000    0.000   -1.137    1.137    0.005
    3 (Z0)         11      23    1    0.011    0.003 1810.259 1812.532   90.760
    4 gamma         1      22    1   -0.012   -0.002    5.371    5.371    0.000
    5 gamma         1      22    1    0.000    0.000 1683.149 1683.149    0.000
    6 gamma         1      22    1    0.000    0.000-3498.863 3498.863    0.000
    7 mu-           1      13    3   12.468  -25.466 1612.743 1612.992    0.106
    8 mu+           1     -13    3  -12.457   25.469  197.516  199.540    0.106
                   sum:  0.00        -0.001    0.001   -0.084 6999.916 6999.916


 ***************************************************************************
 *                       KK2f_Finalize  printouts                          *
 *    7000.00000000                 cms energy total         cmsene     a0 *
 *             5000                 total no of events       nevgen     a1 *
 *               ** principal info on x-section **                         *
 *     233.95163953  +- 1.04896414  xs_tot MC R-units        xsmc       a1 *
 *       0.41468908                 xs_tot    picob.         xSecPb     a3 *
 *       0.00185933                 error     picob.         xErrPb     a4 *
 *       0.00448368                 relative error           erel       a5 *
 *       0.82048782                 WTsup, largest WT        WTsup     a10 *
 *                       ** some auxiliary info **                         *
 *       0.00219522                 xs_born   picobarns       xborn    a11 *
 *       0.73760000                 Raw phot. multipl.                 === *
 *       5.00000000                 Highest phot. mult.                === *
 *                         End of KK2f  Finalize                           *
 ***************************************************************************

\end{verbatim}
}
\end{widetext}


\end{document}